\documentclass[journal,draftclsnofoot,onecolumn,12pt]{IEEEtran}
\usepackage{amsmath,amssymb,amsfonts,epsfig,amsthm,etoolbox,epstopdf}
\usepackage[caption=false,font=footnotesize]{subfig}
\usepackage{dsfont,algorithmic} 
\usepackage[mathscr]{euscript}
\usepackage{enumitem,afterpage}

\usepackage[square, comma, numbers, sort&compress]{natbib}
\usepackage{multirow}
\usepackage[dvipsnames]{xcolor}

\usepackage{tikz,pgfplots}
\usetikzlibrary{positioning,decorations.pathreplacing}
\usetikzlibrary{fit}
\usetikzlibrary{patterns}
\usetikzlibrary{shapes,snakes}
\usetikzlibrary{arrows, calc}
\usetikzlibrary{plotmarks}

\pgfplotsset{compat = 1.8}

\theoremstyle{definition}

\theoremstyle{proof}

\title{Availability-Aware Cell Association for Hybrid Power Supply Networks with Adaptive Bias}

\author{Fanny Parzysz, \IEEEmembership{Member, IEEE,
} and Christos Verikoukis, \IEEEmembership{Senior Member, IEEE}
{\color{gray}\thanks{Fanny Parzysz was with the Electronics Department, University of Barcelona, Spain (e-mail: fanny.parzysz@ieee.org).
Christos Verikoukis is with the Telecommunications Technological Centre of Catalonia (CTTC), Castelldefels, Spain (e-mail: cveri@cttc.es).
}}}

\begin{document}

\maketitle

\begin{abstract}
New challenges have emerged from the integration of renewable energy sources within the conventional electrical grid which powers base stations (BS). Energy-aware traffic offloading brings a promising solution to maintain the user performance while reducing the carbon footprint.
Focusing on downlink cellular networks consisting of on-grid, off-grid and hybrid BSs, we propose a novel power-aware biased cell association 
where each user independently partitions BSs into two sets and applies different association biases for each, depending on the type of power, renewable or not, that can be requested for service. The gain provided by such strategy regarding the probability of power outage and the average grid power consumption is investigated.
To capture their dual nature, the bias applied for association with a hybrid BS is not constant among users nor over time, and is dynamically tailored to the fluctuations of the BS battery level, the user power requirement and the estimated power consumed to serve other users potentially associated with the same BS. Such approach allows to efficiently share the available energy among BSs and turns high heterogeneity in the BS powering into advantage.
\end{abstract}

\begin{IEEEkeywords}
Downlink cellular networks, user association, energy harvesting, bias optimization, power availability, energy consumption, stochastic geometry.
\end{IEEEkeywords}
\section{Introduction}


The next-generation (5G) systems are expected to serve an unprecedented number of devices, estimated at over six connected terminals per person \cite{Buzzi2016}, and to largely rely on small-cell base stations (BS), i.e. low-powered radio access nodes having a coverage range from some tens to some hundreds meters.
The fast escalation of the energy demand, together with increasing environmental concerns on greenhouse gas emissions, has urged the integration of renewable energy sources \cite{Farhangi2010,Fang2012, Mao2015,huawei}, resulting in networks with heterogeneous power supply, composed of (i) on-grid BS, solely connected to the conventional carbon-based power grid, (ii) off-grid BSs, solely relying on energy harvesting and (iii) hybrid BSs, combining both types of power supply. 

Traffic offloading was originally proposed to boost the network capacity, by leveraging under-used spectrum across the different existing radio access technologies \cite{Huang2015,Han2010cellular} and efficiently re-routing the user traffic towards available pico-BSs, femto-BSs or WiFi hot spots. But it also plays a fundamental role in green wireless cellular networks.
A wide range of energy-aware offloading strategies has been proposed for on-grid networks \cite{Han2014,Singh2013,Chen2015}.
However, such conventional approaches do not encompass the unique challenges raised by energy harvesting \cite{Zhang2016}.
Firstly, renewable energy is generally unreliable and unpredictable, depending on the weather condition. Secondly, the traffic load may not be distributed in accordance to the energy harvesting random process, leading to energy waste in parts of the network and energy shortage in other parts.
Thirdly, the coexistence of different power supplies within the same network, each having its own operating characteristics, calls for a refined management of the available energy.

Optimizing the user association issue in small-cells networks with heterogeneous power supply has been investigated in the literature with the objective of minimizing the grid energy consumption, e.g. in \cite{Liu2015,Liu2016,Zhang2016}, maximizing the data rate \cite{Rubio2014,Song2014}, balancing the trade-off existing between traffic delay and power consumption \cite{Liu2014} or minimizing the outage probability due to energy shortage \cite{Wang2015}. 
The proposed optimized decisions are processed at a central (possibly virtual) node or are implemented in a distributed manner through iterative algorithms. Although in both cases, optimality is reached only through extensive causal (possibly non-causal) information exchange on the instantaneous channel and battery states, the extra cost of such additional signaling is generally ignored in the evaluation of the overall energy performance and thus, questions the gains provided by centralized strategies.

This motivates us to rather focus on distributed one-shot cell association schemes, where each user associates with a BS independently of others, using local CSI/BSI only, and in a self-organized manner. But the resulted information shortage, jointly with the uncertainty in the energy arrivals and battery limitations, (i) lets the possibility for users to associate with a BS that cannot guarantee service due to low battery level and (ii) may result in poor energy management, by excessively draining the BS batteries at each time slot or conversely, by underusing the available renewable energy.
While this first issue has been addressed in our previous work \cite{EHBS_Journal1}, we here propose to regulate the energy consumption from both grid and renewable supplies by the use of association weights, or biases, that are applied to artificially squeeze or extend the BS coverage.

A general framework has been proposed in \cite{Sakr2014} to assess the impact of biases on the performance of multi-tier networks with RF energy harvesting, and several analyses have been provided, notably in \cite{Song2014, Yang2016} for throughput optimization and outage probability respectively. 
But the strategies proposed in these works consider association biases which only depend on the BS tier, irrespective of the BS current battery level or the user power requirement.
The use of such constant biases cannot capture the dual nature of hybrid base stations, since it does not differentiate in this case the consumed carbon-based grid energy and the renewable one, harvested from the environment. Dynamically adapting the association biases to the instantaneous energy resource and user requirement opens perspectives on green traffic offloading that have not been explored so far.


\textit{Contributions:} In this paper, we aim to minimize the power consumed from non-renewable sources while meeting a received power constraint at any served users, by utilizing the power-aware cell association of \cite{EHBS_Journal1} and proposing for it a novel strategy based on adaptive biases.
To our best knowledge, this work is the first to propose and analyze non-constant adaptive biases for user association.

Different from existing schemes, our approach is user-centric and only requires periodical broadcast of the BS battery levels. 
Each user computes, independently from others, the energy required by neighboring BSs to satisfy a received power constraint over one time slot and then, partitions BSs depending on whether or not it can be served from renewable energy.
Next, biasing is applied for association decision to prompt users to be served by BSs having sufficient renewable green power.
In this work, we propose a novel approach where biases are dynamically tailored, \emph{at each user and time slot}, to the energy stored in the BS battery, the user required power, accounting for path-loss and shadowing, but also the estimated power consumed to serve other users potentially associated with the same BS.

Next, adjusting the association biases allows to regulate the average amount of energy stored in the BS battery, and thus to control the probability of power outage and the grid power consumption. 
Analyzing such trade-off is the second objective of this work.
 As most of small-cell infrastructures are opportunistically deployed, resulting in irregularly-shaped networks, modeling the node position as random variables allows to analyze the network performance using tools of stochastic geometry and alleviates the need for extensive time-consuming Monte-Carlo simulations. We thus present a comprehensive analysis for the proposed strategy and derive closed-form expression for the performance gain.
 Finally, we show that dynamically associating with BSs depending on the current battery level achieves significant power saving while, at the same time, taking advantage of higher network heterogeneity.

The paper is organized as follows. 
The network model and main assumptions are described in Section \ref{sec:system_model}. The proposed cell association  with adaptive biases is presented in Section \ref{sec:def_cell_asso} and its performance is analyzed in Section \ref{sec:performance_analysis}. Numerical simulations are presented in Section \ref{sec:simulations} and Section \ref{sec:conclusion} concludes this paper.

\section{System Model and Assumptions}
\label{sec:system_model}

We describe in this section the network model, the assumptions for energy harvesting, together with the battery model at off-grid and hybrid base stations.  Table \ref{table:notation} summarizes the notation used in this paper.

\begin{table*}

\caption{Considered notations}
\label{table:notation}
\centering
\hspace*{-10pt}\begin{tabular}{|c|l|}
\hline
 $k / j $ & Considered BS / User\\
  $l / l^{\star} $ & Power buffer state / Lowest buffer state providing availability
\\
\hline 
 $\mathcal{B}_{X}$ / $\lambda_{X}$ 		& X-BS position in the network (PPP / density) \\[3pt]
  $\mathcal{A}_{j}^{(X)}$ / $\Lambda_{X}^{(A)}$ 		& Subset of Available X-BSs, serving user $j$ from renewable energy
  \\[3pt]
   $\mathcal{G}_{j}^{(X)}$ / $\Lambda_{X}^{(G)}$ 		& Subset of Non-available X-BSs, serving user $j$ from the grid supply
   \\[3pt]
   $\mathbb{P}_{\text{X}}^{\left(\mathcal{A}\right)} (p \; \vert  \; l)$ / $\mathbb{P}_{\text{X}}^{\left(\mathcal{G}\right)} (p \; \vert  \; l)$		& Probability to associate with a X-BS and request green/grid power, given a battery state $l$
   \\[3pt]

  $\mathcal{U}$ / $\omega$ 		& User position in the network (PPP / density)
  \\[3pt]
   $\Omega_{\text{X}}^{(\mathcal{A})}$ / $\Omega_{\text{X}}^{(\mathcal{G})}$		& Density of the users served by a X-BS from renewable source/grid supply \\[3pt]
\hline

 $ \mathsf{P}_{\mathsf{Rx}}$ / $\mathsf{P}_\mathsf{Tx}^{(\max)}$					& Received power constraint / Maximum power consumed from the grid supply
 \\[3pt]
  $p_{kj}$		& Power required to send data from BS$_k$ to user $j$
  \\[3pt]
  $ \mathbb{P}_{\text{T}}^{(\mathsf{A-X})}(m \; \vert \; l)$ / $ \mathbb{P}_{\text{T}}^{(\mathsf{G-X})}(m \; \vert \; l)$ & Probability that a X-BS consumes exactly $m$ power units from the battery / grid
  \\[3pt]
   $ \mathsf{P}_{\mathsf{X}}^{(G)}$ &  Average power consumed from the grid at a X-BS
\\[3pt]
$\mathbf{v}^\mathsf{(X)}$ / $\mathbf{P}^\mathsf{(X)}$ & Probability vector and transition matrix of the battery states at a X-BS \\[3pt]
\hline
\end{tabular}
\end{table*}

\subsection{A PPP-based network}

We consider a downlink cellular network which consists, as in \cite{Zhang2016}, of (i) on-grid BSs (OG-BS), connected to the power grid, (ii) energy-harvesting BSs (EH-BS), solely powered by energy scavenging and (iii) hybrid base stations (HY-BS) which are both connected to the power grid and provided with energy harvesting facilities. 
The BSs of type $\mathsf{X} $, with $\mathsf{X} \in \left\lbrace \text{OG, EH, HY} \right\rbrace$, are distributed according to an independent homogeneous PPP $\mathcal{B}_X$, with density $\lambda_X$.

Next, users are distributed according to an independent homogeneous PPP $\mathcal{U}$, with density $\omega$, and are assumed in coverage of more than one BS.
Served users are separated in time, frequency or both (OFDMA), implying that there is no intra-cell interference, as in \cite{Dhillon2014_Fundamentals}. We further assume a power-limited framework, where the available time-frequency resource is large enough to accommodate any user requiring service, as long as sufficient power is available. Such baseline assumption provides a tractable benchmark for performance analysis and is supported by the densification of next-generation cellular networks, where BSs serve only a few users each.

\subsection{Channel attenuation and Required power}
\label{sec:channel_model}

The channel model accounts for path-loss and shadowing. All links are assumed mutually independent and identically distributed. 
First, the path-loss from BS$_k$ to user U$_j$ is given by $\kappa r_{k,j}^{\alpha}$, where $r_{k,j}$ is the distance between them, $\kappa$ is the free-space path-loss at a distance of 1m and $\alpha$ is the path-loss exponent.
Second, the shadowing attenuation $\chi_{k,j}$ from BS$_k$ to U$_j$  follows a log-normal distribution, with zero-mean and standard deviation $\sigma$. Since fast fading can be hardly tracked for cell association, it is not taken in to account.

As in \cite{EHBS_Journal1}, we consider a power allocation which minimizes the transmit power required to satisfy a received power constraint $\mathsf{P}_{\mathsf{Rx}}$. More commonly envisaged for uplink transmissions, such a model allows more efficient use of the renewable energy available in the battery \cite{Liu2015, Mao2015}. A BS with low power in its battery can nevertheless serve nearby users, with low power requirement. $p_{kj}$ is defined as the power requested by user $j$ to receive data from BS$_k$, i.e.
\begin{align}
p_{kj} = \mathsf{P}_{\mathsf{Rx}} \; \frac{\kappa r_{k,j}^{\alpha}}{\chi_{k,j}}.
\label{eq:p_kj}
\end{align}
%
While the required power $p_{kj}$ is limited by the battery level at EH-BSs, we assume a power constraint $\mathsf{P}_\mathsf{Tx}^{(\max)}$ at both OG-BSs and HY-BSs. If  $p_{kj} > \mathsf{P}_\mathsf{Tx}^{(\max)}$, user $j$ is declared in outage with respect to BS$_k$.

\subsection{A Markov chain model for the battery}
\label{sec:model_buffer}

We assume a slotted-time model, with a slot duration of $\tau = 1$.
The amount of power\footnote{The transmission and harvesting processes are considered within the duration of one time slot, such that the analysis proposed in this paper is based on power rather than energy, without loss of generality.}
 that is available in the buffer of EH-BSs and HY-BSs are broadcast towards users at the beginning of a time slot, as part of signaling in control channels.
We assume that the time slots at EH-BSs, OG-BSs and HY-BSs match, and that battery broadcasts are performed simultaneously.
 Based on such information, users perform cell association during this time slot, and are selected for service at the end of it. Effective data transmission occurs in the next time slot and all selected users are served simultaneously. 

Both EH-BSs and HY-BSs are equipped with batteries, or power buffers, potentially with distinct energy harvesting capabilities. We assume that the power stored in a battery is solely issued from renewable sources, and conversely, that all the green power is stored in the battery before consumption. This implies that the power consumed from the grid supply does not affect the battery level.
As in \cite{Sakr2015, EHBS_Journal1}, both the harvested and consumed powers are continuous random variables but are discretized into a finite number of levels for the purpose of analysis. The battery fluctuations, as broadcast at each time slot, depend solely on the previous battery states, the user associations at previous time slot and the power harvested during $\tau$. They are thus modeled by a finite-state Markov chain.
For $\mathsf{X} \in \left\lbrace \text{EH, HY} \right \rbrace$, we respectively define $\varepsilon^\mathsf{(X)}$, $l^\mathsf{{(X)}}$ and $\mathsf{L^{(X)}}$ as the step size (or power unit), the current battery level and the battery capacity of a X-BS (in power units). 
In addition, the probability that the battery has $l$ power units is denoted as $v_l^\mathsf{(X)}$ and the probability vector of the battery states as $\mathbf{v}^\mathsf{(X)} = \left[ v_0^\mathsf{(X)} v_1^\mathsf{(X)} \ldots v_{\mathsf{L}}^\mathsf{(X)} \right]$. 
We also denote as $\mathbb{P}_{\overrightarrow{lq}}^\mathsf{(X)}$ the probability to go from state $l$ to state $q$  from one time slot to the next one at a X-BS. 
The related transition matrix is referred as $\mathbf{P}^\mathsf{(X)} = \left[ \mathbb{P}_{\overrightarrow{lq}}^\mathsf{(X)}\right]$.

Without loss of generality, a general energy harvesting model is considered, where the power arrivals follow a Poisson process of intensity $\lambda_e$. 
Such a model is used for example for solar photo-voltaic panels, but the analysis proposed in this work is valid for other harvesting processes, by considering other function $\mathbb{P}_{\text{H}}$.

Regarding the model for battery depletion, we assume as in \cite{Sakr2015, EHBS_Journal1} that the power $p_{kj}$ consumed to send data from BS$_k$ to user $j$ is rounded up to the nearest battery unit. 
Given that the battery levels are broadcast only periodically, more than one user can associate with the same BS based on the same battery state information.
Therefore, from one time slot to the following one, the battery level is decreased by an amount of power equal to the sum of the powers requested by all users selected during this time slot. 
The probability $\mathbb{P}_{\text{T}}^{(\mathsf{X})}(m \; \vert \; l)$ to consume a \emph{total} of $m$ power units given that X-BS$_k$ has $l$ power units in its battery is equal to
\begin{align}
\mathbb{P}_{\text{T}}^{(\mathsf{X})}(m \; \vert \; l) =  \mathbb{P}\left( \sum_j \left \lceil p_{kj} \right \rceil = m \; \vert \; l \right),
\label{eq:prob_T}
\end{align}
where the summation is over the set of users served by BS$_k$. We assume that associated users are selected for service \textit{in ascending order} of their required transmit power. 
The computation of this probability is one of the major challenges solved in Section \ref{sec:performance_analysis}.

\section{Biased Cell Association with Adaptive Biases}
\label{sec:def_cell_asso}

This section describes the considered cell association with BS partitioning and adaptive biases. The case with EH-BSs only is first described, next extended to the general case. We conclude by discussing on the particular case of HY-BSs.

\subsection{User association for homogeneous networks \cite{EHBS_Journal1}}

We first focus on networks consisting of EH-BSs only, for which a cell association jointly accounting for the BS available power and the user power requirement of Eq. \eqref{eq:p_kj} has been proposed in \cite{EHBS_Journal1}. 
It is performed in three steps, briefly restated in the following:
\begin{enumerate} [label=\textit{$\bullet$ Step \arabic*:} ,align=left]
	\item \emph{Estimating the BS power availability:} A base station BS$_k$ is declared available for user $j$ if the amount of power $l_k\varepsilon$ (broadcast at the beginning of the current time slot) is sufficient to accommodate the requested power $p_{kj}$ in addition to the \textit{estimated} power $\widetilde{\mathsf{P}}_{k\setminus j}$ consumed by other users potentially associated with BS$_k$, i.e.
	\begin{align}
	p_{kj}+ \widetilde{\mathsf{P}}_{k\setminus j}  \leq l_k\varepsilon
	\quad 
		\text{where} \quad & \widetilde{\mathsf{P}}_{k\setminus j} \triangleq  \omega \Upsilon \frac{2 / \alpha} {2 / \alpha +1} \left( p_{kj} \right)^{\frac{2}{\alpha}+1}
		\label{eq:power_availability_criteria}
	\\
	\text{and} \quad & \left \lbrace \begin{array}{l l}
	\Upsilon &= \pi \left(\frac{1}{\mathsf{P_{Rx}} \kappa}\right)^{\frac{2}{\alpha}} \exp \left(  \frac{2/\alpha}{\zeta} \mu + \frac{1}{2} \left(\frac{2/\alpha}{\zeta}\right)^2 \sigma^2 \right) 
	\\
	\zeta &= 10/ \ln(10)
	\end{array} \right. \nonumber
	\end{align}
	The estimate $\widetilde{\mathsf{P}}_{k\setminus j}$ characterizes the BS availability and is computed independently at each user. If overestimated, users will more likely declare BSs as unavailable. If underestimated, too many users may associate with the same BS, leading to power outage. We highlight that a BS may be available for one user and not for another one.
	
	\item \emph{Effective association:} Then, each user effectively associate with the available BS that minimizes the required transmit power. No multi-cell transmission nor clustering is assumed in this work.
	
	\item  \emph{User selection:} Finally, each BS selects the users which can be served among the associated ones. Since users associate with a given BS based on an estimate of the total power consumption, such BS may not be able to serve all associated users
	To maximize the number of served users, each BS selects associated users in ascending order of their power requirement, till all associated users are selected or till the battery is empty.	
	
\end{enumerate}

We refer the reader to \cite{EHBS_Journal1} for further details on the cell association. Such strategy has been shown to provide significant reduction of the probability of power outage, exceeding 50\% in case of bursty energy arrivals, while requiring little extra overhead to broadcast the BS battery levels at each time slot.

\subsection{Application to networks with heterogeneous power supply}
\label{sec:cell_association}

Different from the analysis of \cite{EHBS_Journal1}, we now move onto considering general networks, consisting of EH-BS, HY-BS and OG-BS.

\subsubsection{Step 1: BSs partitioning}

To minimize the power consumed from the conventional grid supply, users are prompted to associate in priority with base stations which can guaranty service solely using the power issued from renewable source, i.e. the power harvested from the environment and stored in the battery.
By applying the power-availability criteria of Eq. \eqref{eq:power_availability_criteria}, each user partitions neighboring X-BSs 
into two sets. 
If the battery level is high enough compared to the required power, a X-BS is declared available and power from renewable source will be requested if user $j$ associates with it. The sub-set of corresponding BSs is denoted by $\mathcal{A}_{j}^{(\mathsf{X})}$, with $\mathcal{A}$ for \emph{available}. 
Otherwise, BSs are declared non-available and power from the grid supply will be requested. This subset of BSs is denoted as $\mathcal{G}_{j}^{(\mathsf{X})}$, with $\mathcal{G}$ for \emph{grid}. We highlight that such partitioning is different from one user to another and jointly depends on the battery level and the power requirement.
Defining $g_{\mathsf{X}} \left(p_{kj}\right) = p_{kj} + \widetilde{\mathsf{P}}_{k\setminus j}^{(\mathsf{X})}$,
\begin{align}
\mathcal{A}_{j}^{(\mathsf{X})} & \triangleq \left \lbrace \text{BS}_k \in  \mathcal{B}_{\mathsf{X}}:  
g_{\mathsf{X}} \left(p_{kj}\right)  \leq l_k\varepsilon^{(\mathsf{X})}  \right \rbrace
\quad \text{and}
\label{eq:set_A_with_gx}
\\
\mathcal{G}_{j}^{(\mathsf{X})} & \triangleq \left \lbrace \text{BS}_k \in  \mathcal{B}_{\mathsf{X}}:  
g_{\mathsf{X}}^{-1} \left(l_k\varepsilon^{(\mathsf{X})} \right)   \leq  
p_{kj} \leq \mathsf{P}_\mathsf{Tx}^{(\max)} \right \rbrace.
\label{eq:set_G_with_gx}
\end{align}

\textit{Remarks:} 
For OG-BSs, $\mathcal{A}_{j}^{(\mathsf{OG})} = \emptyset$ (equivalently HY-BSs with empty battery) and for EH-BSs, $\mathcal{G}_{j}^{(\mathsf{EH})} = \emptyset$ (HY-BSs with $\mathsf{P}_\mathsf{Tx}^{(\max)} = g_{\mathsf{EH}}^{-1} \left(l\varepsilon^{(\mathsf{EH})} \right)$ at each instant).
Any BS which belongs neither to $\mathcal{A}_{j}^{(\mathsf{X})} $ nor to $\mathcal{G}_{j}^{(\mathsf{X})} $ is not considered for cell association.

\subsubsection{Step 2: Effective association}
A user $j$ associates with the base station BS$_{k^{\star}}$ which consumes the less \textit{biased} power to satisfy the received power constraint $\mathsf{P}_{\mathsf{Rx}}$:
\begin{align}
 \left. \begin{array}{l}
k^\dagger  = \underset{k \in \mathcal{A}_{j}} {\arg \min} \left\lbrace p_{kj} \right \rbrace \\
k^\ddagger  = \underset{k \in \mathcal{G}_{j}} {\arg \min} \left\lbrace p_{kj} \right \rbrace
\end{array}
\right \rbrace
\; \text{and} \;  k^{\star} = \underset{k \in  \left \lbrace 
	k^\dagger, k^\dagger
	\right \rbrace} {\arg \min} \left\lbrace \beta_{kj} p_{kj} \right \rbrace
\quad \text{where} \; \beta_{kj} = \left \lbrace \begin{array}{l l}
\beta_A & \text{if} \; k \in \mathcal{A}_{j}
\\
\beta_G & \text{if} \; k \in \mathcal{G}_{j}
\end{array}\right.
\label{eq:bias_definition}
\end{align}

In this, $\beta_{kj}$ denotes for a power bias, or association weight, which is applied for cell association of user $j$ with BS$_k$. 
Such biases allow to control traffic off-loading, from BSs with low battery to BSs with higher battery. 
Together with the association request, each user announces the requested power supply.
If BS$_{k^{\star}} \in \mathcal{A}_{j}$, user $j$ requests renewable energy, whether it is an EH-BS or a HY-BS. Otherwise, it requests grid supply, whether it is an OG-BS or a HY-BS.

\subsubsection{Step 3: User selection} Both EH-BSs and HY-BSs may need to drop users which cannot be served given their effective battery level. While HY-BSs can consume extra grid power to serve such users, power outage is declared at EH-BSs.

\subsection{On the particular case of hybrid BSs}

HY-BSs are both provided with energy harvesting facilities and access to the power grid. While a user can only request renewable power (resp. grid power) if associated with an EH-BS (resp. OG-BS), it is free to choose the power source if associated with a HY-BS. We \textit{do not} assume that grid power is consumed at a HY-BS only if the battery is empty. As discussed in Section \ref{sec:simulations}, this significantly improves the network performance. 

Another consequence is that HY-BSs manage two sets of users, one being served with renewable supply and one with grid supply.
Different from existing literature on biased cell association,
the bias applied by a user for associating with a HY-BS is not constant over the whole set of HY-BSs. It varies from one time slot to the other and from one user to the other, according to the user power requirement and the battery level fluctuations.
As far of our knowledge, this is the first non-constant \emph{adaptive} bias proposed for cell association.

\section{Performance analysis of cell association with adaptive biases}
\label{sec:performance_analysis}

\begin{figure}
	\centering
	\subfloat[$\beta_A = 0.1$]{ \hspace{-18pt} \includegraphics[width=0.4\columnwidth]{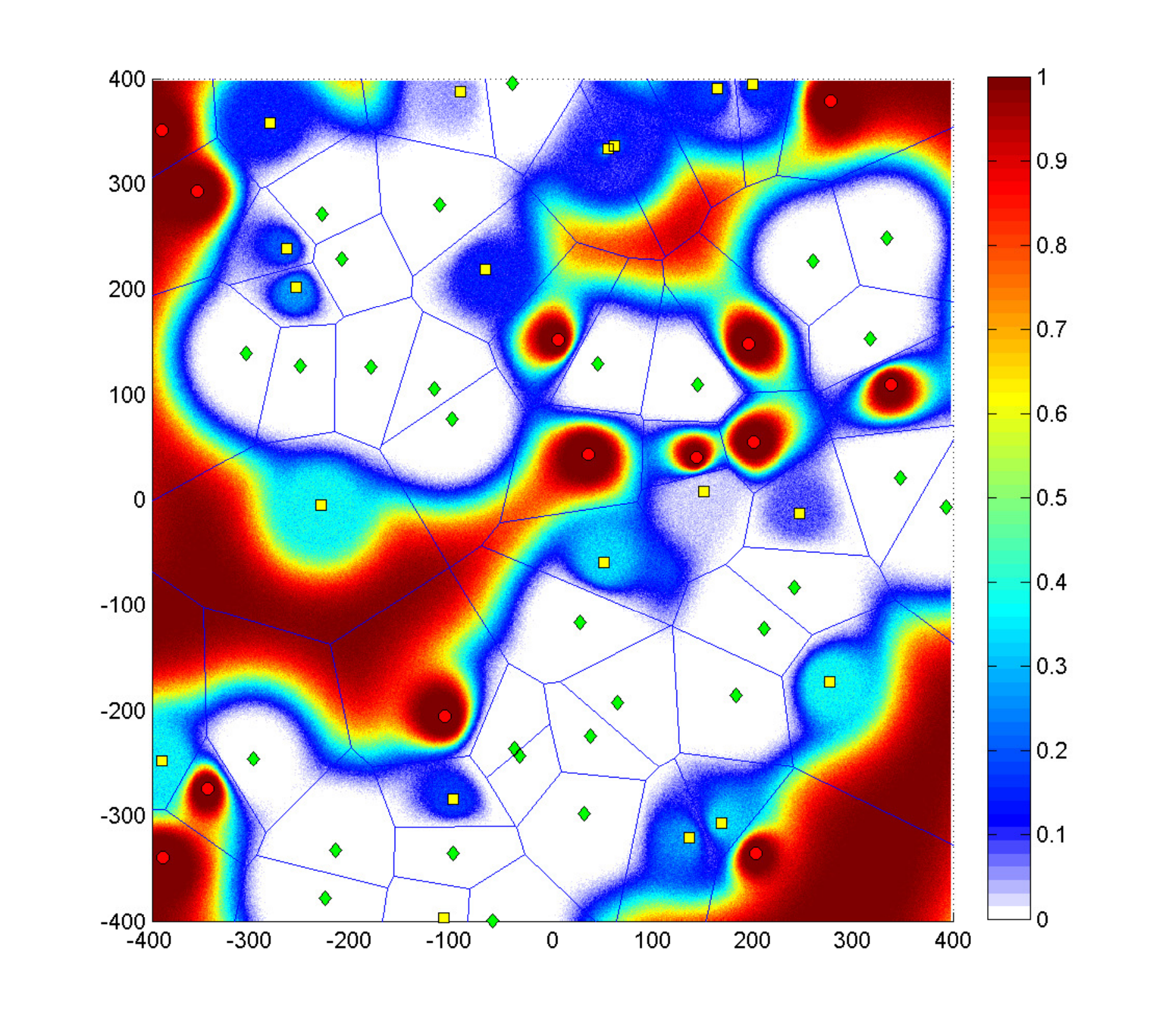}} 
	\subfloat[$\beta_A = 2$]{ \hspace{-10pt}\includegraphics[width=0.4\columnwidth]{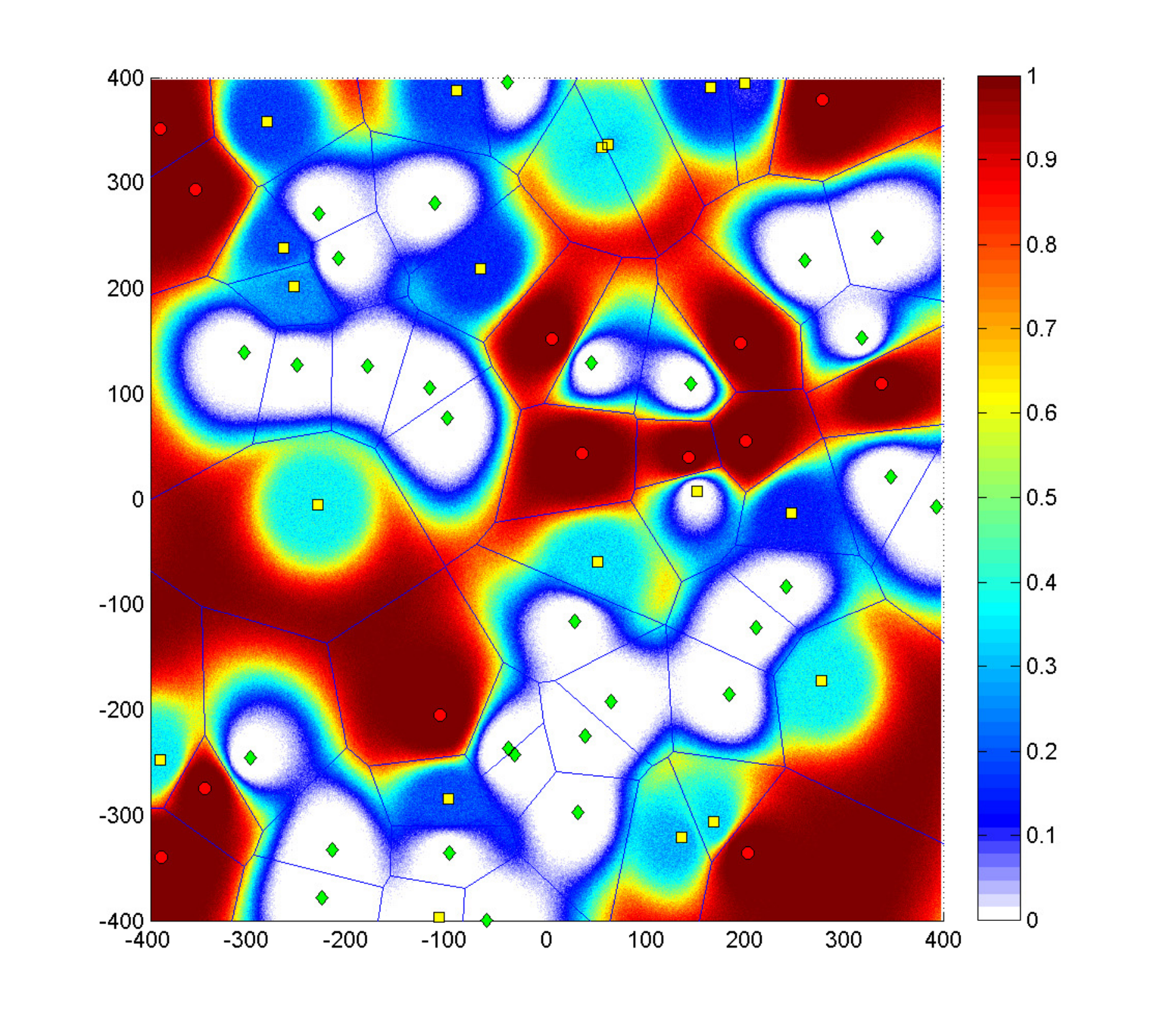}}
	
	\caption{\footnotesize  Probability to be served using power from conventional grid supply ($\omega = \frac{20}{\pi (100)^2}$, $\lambda_e^{(\mathsf{EH})} = 0.1 \mathsf{L}$, $\lambda_e^{(\mathsf{HY})} = 0.05 \mathsf{L}$, $\mathsf{P}_{\mathsf{Rx}} =-60$dBm, $\beta_G = 1$, 
		Red circles: OG-BSs, Yellow squares: HY-BSs, Green diamonds: EH-BSs)
	}
	\label{fig:illustration_power consumption}
\end{figure}

As illustrated in Figure \ref{fig:illustration_power consumption}, the association biases significantly impact the probability for a user to be served using power from the conventional grid supply. Setting $\beta_A$ close to zero forces users to associate with a BS equipped with harvester as soon as energy is available.
On the contrary, by increasing  $\beta_A$, users are more likely associated with HY-BSs or OG-BSs and requesting power from the grid, such that batteries are maintained at a higher level throughout the network.
In this section, we analyze the proposed scheme to allow the quantification of such trade-off.

\subsection{Problem characterization for bias optimization}

Monte-Carlo simulations may rapidly turn out to be excessively time-consuming in regard to the considerable number of samples necessary to average both the node locations and the battery level fluctuations \cite{Andrews2011}. As a consequence, we propose to analyze the performance of the proposed strategy in terms of the probability of power outage $\mathbb{P}_{\text{out}}$ and the average power per area unit, $ \mathsf{P}_{\mathsf{OG}}^{(G)}  + \mathsf{P}_{\mathsf{HY}}^{(G)} $, that is consumed from the grid supply at both OG-BSs and HY-BSs.

At a given time slot, the current battery levels, as broadcast by HY-BSs and EH-BSs, fully determine the BS partitioning into sets $\mathcal{A}$ and $\mathcal{G}$ and give the probability of association to an EH-BS, a HY-BS or an OG-BS. Reciprocally, such probability affects the power consumption, from green or carbon-based power supply, and thus, determines the battery levels at the next time slot. 
Thus, to any pair of biases $(\beta_A, \beta_G)$ corresponds an equilibrium distribution of the battery states at EH-BSs and HY-BSs, and of the number of users served by EH-BSs, HY-BSs and OG-BSs.
Characterizing such equilibrium is essential to compute the performance of the proposed strategy.
While the Markov chain analysis described in Subsection \ref{sec:markov_algo} allows to compute the stationary distributions for $\mathbf{v}^\mathsf{(HY)}$ and $\mathbf{v}^\mathsf{(EH)}$,
the stochastic geometry tools used in Subsection \ref{sec:density} allow to characterize (i) the average number of available and non-available X-BSs per area unit, i.e. the density of sets $\mathcal{A}_0$ and $\mathcal{G}_0$, for a typical user $U_0$, (ii) the probability of association with an EH-BS, a HY-BS or an OG-BS, and (iii) the density of users served by any X-BS, for a given set of biases $(\beta_A, \beta_G )$ and battery probability vector $\mathbf{v}^\mathsf{(HY)}$ and $\mathbf{v}^\mathsf{(EH)}$. 
Performance is deduced in subsection \ref{sec:performance}.

\subsection{On the stationary distribution of the battery states}
\label{sec:markov_algo}

The stationary distribution of $\mathbf{v} = \left[ \mathbf{v}^\mathsf{(EH)} \mathbf{v}^\mathsf{(HY)} \right]$ is numerically solved by adapting the algorithm proposed in \cite{EHBS_Journal1} to networks with heterogeneous power supply. We restate its main steps, as an overview of the analysis proposed in next sub-sections.
Inspired from the fixed-point method, a first guess $\mathbf{v}_{(0)}$ for the solution (e.g. uniform distribution) is considered, then successive approximations of the stationary distribution are computed as $\mathbf{v}_{(i+1)} =\mathbf{v}_{(i)} \mathbf{P}$, where $\mathbf{P} =
\left[\begin{array}{cc}
\mathbf{P}^\mathsf{(EH)} & 0 \\
0 & \mathbf{P}^\mathsf{(HY)}
\end{array}\right]$ and $\mathbf{P}^\mathsf{(X)}$ is the transition matrix of the Markov chain modeling the battery at X-BSs.
The i$^{th}$ iteration consists of:
\begin{enumerate}[label=\textit{$\bullet$ Step \arabic*:} ,align=left]
	
	\item \textit{BS partitioning:} Given $\mathbf{v}_{(i)}$, compute the densities of $\mathcal{A}$ and $\mathcal{G}$, respectively denoted as $\Lambda_{\text{X}}^{(A)}$ and $\Lambda_{\text{X}}^{(G)}$.
	
	\item \textit{Effective association:} Deduce the probability $\mathbb{P}_{\text{X}}^{\left(\mathcal{A}\right)} (p \; \vert  \; l)$ (resp. $\mathbb{P}_{\text{X}}^{\left(\mathcal{G}\right)} (p \; \vert  \; l)$) to associate with an EH-BS or a HY-BS (resp. an OG-BS or a HY-BS) and to request renewable power (resp. carbon-based power), for any battery state $l$. 
	
	\item \textit{User selection:} Compute the density $\Omega_{\text{X}}^{(\mathcal{A})} (p \; \vert \; l)$ (resp. $\Omega_{\text{X}}^{(\mathcal{G})} (p \; \vert \; l)$) of users served by a X-BS using power from the green (resp. grid) supply, for any state $l$. 
	
	\item \textit{Power consumption:} Using Theorem 1 of \cite{EHBS_Journal1}, deduce from the density $\Omega_{\text{X}}^{(\mathcal{A})} (p \; \vert \; l)$  the probability $ \mathbb{P}_{\text{T}}^{(\mathsf{A-X})}(m \; \vert \; l)$ that a given X-BS consumes exactly $m$ power units from the battery. Similarly, compute the probability $\mathbb{P}_{\text{T}}^{(\mathsf{G-X})}(m \; \vert \; l)$ to consume $m$ units from the power grid. Then, derive the new transition matrix $\mathbf{P}$ and vector $\mathbf{v}_{(i+1)}$ as in \cite{EHBS_Journal1}.
	
\end{enumerate}

In this algorithm, steps 1 and 2 characterizes the network from the point of view of the typical user, while steps 3 and 4 move onto the perspective of a base station. The equilibrium distributions of the battery states at EH-BSs and HY-BSs are \emph{not} independent and have to be \emph{jointly} computed.
In the following, we extend the analysis proposed in \cite{EHBS_Journal1} for step 1 to networks with heterogeneous supply. Then, we fully analyze steps 2 and 3 which include the novel association biases proposed in this paper. For the benefit of the reader, closed-form expressions required for step 4 and derived in \cite{EHBS_Journal1} are summarized in the Appendix.


\subsection{Characterization of the set of served users}
\label{sec:density}

We consider the i$^{th}$ iteration of the solution algorithm. The biases are set to $( \beta_A, \beta_G )$ and the probability vectors of the BS battery state are equal to $\mathbf{v}^\mathsf{(HY)}_{(i)}$ and $\mathbf{v}^\mathsf{(EH)}_{(i)}$. We drop the index $_{(i)}$ for more readability and start the analysis by two definitions.

First, the power coverage $p_{l}^{(\text{cov,X})}$ is defined as the maximum power issued from renewable sources that can be requested by a user served by a X-BS (EH-BS or HY-BS) having $l$ power units in its battery. For any $U_j$ in coverage of X-BS$_k$,
\begin{align}
p_{kj} \leq \; p_{l_k}^{(\text{cov,X})}
\; \triangleq \;
{g_\mathsf{X}}^{-1} \left(l_k\varepsilon^{(\mathsf{X})} \right)
\label{eq:p_l_cov}
\end{align}
with $g_{\mathsf{X}}$ as given in Eq. \eqref{eq:set_A_with_gx}-\eqref{eq:set_G_with_gx}.
$p_{l}^{(\text{cov,X})}$ depends on the received power constraint $\mathsf{P}_{\mathsf{Rx}}$ and on the energy harvesting rate, implying that $p_{l}^{(\text{cov,EH})}$ is generally not equal to $p_{l}^{(\text{cov,HY})}$. 
Second, we define $l^{\star}$ as the lowest buffer state providing availability of X-BS$_k$. For U$_j$ requiring $p_{kj}$, 
\begin{align}
l^{\star}-1 < g_{\mathsf{X}}(p_{kj}) \leq l^{\star} \quad \Leftrightarrow \quad
p_{l^{\star}-1}^{(\text{cov,X})} < p_{kj} \leq p_{l^{\star}}^{(\text{cov,X})} 
\label{eq:l_star}
\end{align}

\subsubsection{Density of $\mathcal{A}_0$ and $\mathcal{G}_0$ (Step 1 of the algorithm)}

From the properties of displaced PPPs, the point process modeling the powers $p_{k0}$ $\forall k$ 
is distributed according to a non-homogeneous PPP on $\mathbb{R}^+$ of density $\Lambda_{\mathsf{X}} (p) = \lambda_{\mathsf{X}} \Upsilon \left( \mathsf{P}_{\mathsf{Rx}} \right)^{-\frac{2}{\alpha}} p^{\frac{2}{\alpha}}$.
Next, retaining out of this PPP only the BSs which are available for $U_0$ is an independent thinning, whose properties allow to compute the density of $\mathcal{A}_0^{(\mathsf{X})}$.
The retention probability that a X-BS (EH- or HY-BS) has enough available power to serve $U_0$  using renewable supply is equal to $\underset{l = l^{\star}}{\overset{\mathsf{L}}{\sum}} v_l^{(\text{X})}$, such that the density $\Lambda_{\text{X}}^{(A)}$ of $\mathcal{A}_0^{(\mathsf{X})}$ is computed as in Eq. (14) of \cite{EHBS_Journal1}:
\begin{align}
\Lambda_{X}^{(A)} (p) = \underset{l=0}{\overset{l^{\star}-1}{\sum}} v_l^{(X)} \Lambda_{X} (p_{l}^{(\text{cov})}) +  \underset{l = l^{\star} }{\overset{\mathsf{L}}{\sum}} v_{l}^{(X)} \Lambda_{X} (p),
\label{eq:Lambda_B_A}
\end{align}
with $p_{l}^{(\text{cov})}$ as given in Eq. \eqref{eq:p_l_cov}.

Now considering a network with heterogeneous power supply, we deduce from Eq. \eqref{eq:Lambda_B_A} the density  of $\mathcal{G}_{0}^{(\mathsf{X})}$, i.e. the set of X-BSs (HY-BSs or OG-BSs) for which grid power is required:
\begin{align}
\Lambda_{\text{X}}^{(G)} (p)  =
\left \lbrace \begin{array}{ll}
\Lambda_{\mathsf{X}} (p) - \Lambda_{\text{X}}^{(A)} (p) 
 & \text{if } p \leq \mathsf{P}_\mathsf{Tx}^{(\max)}
\\
 \Lambda_{\mathsf{X}} (p) - \Lambda_{\text{X}}^{(A)} (\mathsf{P}_\mathsf{Tx}^{(\max)}) 
 & \text{otherwise}
\end{array}
\right.
\end{align} 

In particular, for EH-BSs, $\Lambda_{\text{EH}}^{(G)} (p) = 0$, $\forall p$ and, for OG-BSs, $\Lambda_{\text{OG}}^{(G)} (p)  = \Lambda_{\mathsf{OG}} (p)$,  $\forall p \leq \mathsf{P}_\mathsf{Tx}^{(\max)}$.

\subsubsection{Probability of association with an X-BS (Step 2 of the algorithm)}

We move onto computing the probability that $U_0$ associates with a base station of type $\mathsf{X} \in \left\lbrace \text{OG, EH, HY} \right\rbrace$. 
As described in Step 2 of Subsection \ref{sec:cell_association}, U$_0$ associates with X-BS$_{0^{\star}}$ if the power $p^{\star}$ required to satisfy the received power constraint $\mathsf{P_{Rx}}$, weighted by the correct bias $\beta^{\star}$, is minimal among all other BSs.

First, let assume that X-BS$_{0^{\star}} \in \mathcal{G}_{0}^{(\mathsf{X})}$, i.e. X-BS$_{0^{\star}}$ consumes grid power to serve U$_0$, implying that $\beta^{\star} =\beta_G$.
Leveraging the reduced Palm distribution \cite{book_sto_geo}, $\forall Y \in \left \lbrace \mathsf{EH}, \mathsf{OG}, \mathsf{HY}\right \rbrace$, the set of Y-BSs (excluding X-BS$_{0^{\star}}$ if $Y=$X) has the same properties as $\mathcal{B}_{\mathsf{Y}}$  and can be partitioned into the subsets 
$\mathcal{A}_{0}^{(\mathsf{Y})}$, for which the association bias $\beta_A$ is applied, and $\mathcal{G}_{0}^{(\mathsf{Y})}$, for which the bias $\beta_G$ is applied.
Therefore, the typical user U$_0$ associates with X-BS$_{0^{\star}}$
\begin{align}
\text{if} \; \forall \mathsf{Y} \; \left\lbrace \begin{array}{l}
 \forall k \in \mathcal{A}_{0}^{(\mathsf{Y})}, \;
 \beta_G p^{\star} <  \beta_A p_{k0}
\; \Leftrightarrow  \; p^{\star} <  \frac{\beta_A}{\beta_G} p_{k0} 
 \\
 \forall k \in \mathcal{G}_{0}^{(\mathsf{Y})}, \;
 \beta_G p^{\star} <  \beta_G p_{k0}
\; \Leftrightarrow  \;   p^{\star} < p_{k0}
\end{array}
\right.
\label{eq:partitioning_YBS}
\end{align}

We now denote $\mathcal{G}$ as the point process modeling such scaled power:
\begin{align}
\mathcal{G} = \underset{Y \in \left \lbrace \mathsf{EH}, \mathsf{OG}, \mathsf{HY}\right \rbrace}{\bigcup} \left \lbrace 
t_k = \frac{\beta_A}{\beta_G} p_{k0} \; : \; k \in \mathcal{A}_{0}^{(\mathsf{Y})}
\right \rbrace
\cup\left \lbrace 
t_k = p_{k0} \; : \; k \in \mathcal{G}_{0}^{(\mathsf{Y})}
\right \rbrace.
\end{align}
Given that it is also a PPP, the void probability of PPP allow to compute the probability that U$_0$ associates with X-BS$_{0^{\star}}  \in \mathcal{G}_{0}^{(\mathsf{X})}$:
\begin{align}
&\mathbb{P}_{\text{X}}^{\left(\mathcal{G}\right)} (p^{\star} \; \vert \; l)  = 
\left \lbrace \begin{array}{ll}
 \exp \left( - \Lambda^{\left(\mathcal{G}\right)} (p^{\star}) \right) & p^{\star} > p_{l}^{(\text{cov,X})}
 \\
 0 & \text{otherwise}
\end{array} \right.
\end{align}
where $l$ is the battery level at X-BS$_{0^{\star}}$ and $\Lambda^{\left(\mathcal{G}\right)}$ the density of $\mathcal{G}$. In particular, $\mathbb{P}_{\mathsf{EH}}^{\left(\mathcal{G}\right)} (p^{\star} \; \vert  l) = 0, \forall  l$.

We now compute $\Lambda^{\left(\mathcal{G}\right)}$. For a BS$_k$ in $\mathcal{A}_{0}^{(\mathsf{Y})}$, we denoting $l^\dagger$ the lowest buffer state providing availability for a user U$_j$ requiring $p_{kj} = t_k \frac{\beta_G}{\beta_A}$, i.e.
\begin{align}
p_{l^{\dagger}-1}^{(\text{cov,X})} < t_k \frac{\beta_G}{\beta_A} \leq p_{l^{\dagger}}^{(\text{cov,X})} .
\end{align}
Once again leveraging the properties of displaced PPP, the process modeling the  $\left \lbrace t_k \right \rbrace$ forms a PPP on $\mathbb{R}^+$ and its density is computed in a similar way as $\Lambda_{\text{X}}^{(A)}$ (Eq.(14) of \cite{EHBS_Journal1}):
\begin{align}
\Lambda_{\text{Y}}^{(\beta A)} \left( t \right) 
= 
\underset{l=0}{\overset{l^{\dagger}-1}{\sum}} v_l^{(\mathsf{X})} \Lambda_{\mathsf{X}} (p_{l}^{(\text{cov,X})}) +  \underset{l = l^{\dagger} }{\overset{\mathsf{L}}{\sum}} v_{l}^{(\mathsf{X})} \Lambda_{\mathsf{X}} \left(\frac{\beta_G}{\beta_A} t\right).
\label{eq:Lambda_betaA}
\end{align}
Note that $l^\dagger \neq l^\star$ if $\beta_A \neq \beta_G$. Therefore, the density of $\mathcal{G}$ is expressed as:
 \begin{align}
 \Lambda^{\left(\mathcal{G}\right)} (p)  =
\Lambda_{\text{EH}}^{(\beta A)} \left( p \right) 
+
\Lambda_{\text{HY}}^{(\beta A)} \left( p \right) 
+
\Lambda_{\text{HY}}^{(G)} \left( p \right) 
+
\Lambda_{\text{OG}}^{(G)} \left( p \right) 
\label{eq:Lambda_matcalG}
 \end{align}
 In this, $\Lambda_{\text{EH}}^{(\beta A)} \left( p \right) + \Lambda_{\text{HY}}^{(\beta A)} \left( p \right) $ corresponds to the base stations (EH- or HY-BSs) which are declared available for U$_0$ (bias $\beta_A$ is applied). $\Lambda_{\text{HY}}^{(G)} \left( p \right) + \Lambda_{\text{OG}}^{(G)} \left( p \right) $ refers to base stations (OG- or HY-BSs) which are declared non-available (bias $\beta_G$ is applied).

A similar analysis can be conducted to compute the probability that U$_0$ associates with X-BS$_{0^{\star}}  \in \mathcal{A}_{0}^{(\mathsf{X})}$, i.e. using bias $\beta_A$.
Focusing on U$_0$,  the power requirements $p_{k0}$ of all other Y-BSs for which the same bias $\beta_A$ is applied are distributed according to a PPP of density $\Lambda_{\text{EH}}^{(A)} \left( p \right) +\Lambda_{\text{HY}}^{(A)} \left( p \right) $, while the scaled power requirements $\frac{\beta_G}{\beta_A} p_{k0}$ of all other Y-BSs (for which bias $\beta_G$ is applied) are modeled according to a PPP with density
\begin{align}
\Lambda_{\text{X}}^{(\beta G)} (p)= \Lambda_{\text{X}}\left( p \frac{\beta_A}{\beta_G}\right) - \Lambda_{\text{HY}}^{(\beta A)} (p),
\end{align}
with $\Lambda_{\text{HY}}^{(\beta A)}$ computed as in in Eq. \eqref{eq:Lambda_betaA} but exchanging $\beta_A$ and $\beta_G$.
Denoting $l$ the battery level at X-BS$_{0^{\star}}$, the probability that U$_0$ associates with X-BS$_{0^{\star}}  \in \mathcal{A}_{0}^{(\mathsf{X})}$ is s.t.:
\begin{align*}
&\mathbb{P}_{\text{X}}^{\left(\mathcal{A}\right)} (p^{\star} \; \vert  \; l)  = 
\left \lbrace \begin{array}{ll}
 \exp \left( - \Lambda^{\left(\mathcal{A}\right)} (p^{\star}) \right) & p^{\star} \leq p_{l}^{(\text{cov,X})}
 \\
 0 & \text{otherwise}
\end{array} \right.
\\
\text{with} \quad &\Lambda^{\left(\mathcal{A}\right)} (p)  =
\Lambda_{\text{EH}}^{(A)} \left( p \right) 
+
\Lambda_{\text{HY}}^{(A)} \left( p \right) 
+
\Lambda_{\text{HY}}^{(\beta G)} \left( p \right) 
+
\Lambda_{\text{OG}}^{(G)} \left( p \frac{\beta_A}{\beta_G}\right)
\end{align*} 
In particular, $\mathbb{P}_{\mathsf{OG}}^{\left(\mathcal{A}\right)} (p^{\star}) = 0$.

\subsubsection{Density of users served by an X-BS (Step 3 of the algorithm)}

So far, the analysis has targeted the typical user's perspective. To characterize the point process of users served by a given X-BS and compute the overall consumed power, we move onto the BS perspective and consider a typical base station of type $X$, denoted as X-BS$_0$. The point process modeling the powers $p_{0j}, \forall j$ required by surrounding users follows a non-homogeneous PPP on $\mathbb{R}^+$ of density $\Omega (p) =  \omega \Upsilon \left( \mathsf{P}_{\mathsf{Rx}} \right)^{-\frac{2}{\alpha}} p^{\frac{2}{\alpha}}$.

\emph{Remark:} Following Step 3 of the cell association procedure, each base station selects the users that can be served among the ones which are associated with it and in ascending order of their biased power requirement, given the battery level limitation. As the power availability criteria is based on an estimate, there exists a non-zero probability that too many users are associated, given the current battery level, and have to be dropped (at EH-BSs) or served using grid power (at HY-BSs). Simulations show that the probability that too many users are associated regarding the battery level is negligible. We thus approximate the set of users selected by any EH-BS or HY-BS by the set of users associated with it.

Retaining out of the PPP modeling the powers $p_{0j}, \forall j$ only the users associated to X-BS$_0$ corresponds to an independent thinning, with retention probability $\mathbb{P}_{\text{X}}^{\left(\mathcal{A}\right)} (p^{\star} \; \vert  \; l)$ if X-BS$_0$ $ \in \mathcal{A}_{j}^{(\mathsf{X})}$ and $\mathbb{P}_{\text{X}}^{\left(\mathcal{G}\right)} (p^{\star} \; \vert  \; l)$ if X-BS$_0$ $ \in \mathcal{G}_{j}^{(\mathsf{X})}$, as computed in previous subsection.
Denoting $l$ the battery level, the set of users which are served from renewable energy has a density equal to 
\begin{align}
&\Omega_{\text{X}}^{(\mathcal{A})} (p \; \vert \; l) = \int_0^p d\Omega(t)
\mathbb{P}_{\text{X}}^{\left(\mathcal{A}\right)} (t \; \vert \; l) 
\mathds{1} \left[t \leq p_{l}^{(\text{cov,X})} \right]
dt
\label{eq:Omega_A}
\end{align}
Similarly, the set of users which are served from the grid supply forms PPP on $\mathbb{R}^+$, with density:
\begin{align}
&\Omega_{\text{X}}^{(\mathcal{G})} (p \; \vert \; l) = \int_0^p d\Omega(t)
\mathbb{P}_{\text{X}}^{\left(\mathcal{G}\right)} (t \; \vert \; l) 
\mathds{1} \left[p_{l}^{(\text{cov,X})}  < t \leq \mathsf{P}_\mathsf{Tx}^{(\max)} \right]
dt
\label{eq:Omega_G}
\end{align}
We highlight that the densities of the served users jointly depend on both probability vectors $\mathbf{v}^\mathsf{(HY)}$ and $\mathbf{v}^\mathsf{(EH)}$ via $\mathbb{P}_{\text{X}}^{\left(\mathcal{A}\right)}$ and $\mathbb{P}_{\text{X}}^{\left(\mathcal{G}\right)}$, whether it is a EH-BS, a HY-BS or a OG-BS. 
Such result concludes the analysis necessary to run the algorithm proposed in Section \ref{sec:markov_algo}. 

\subsection{Computation of considered performance metrics}
\label{sec:performance}

Once the stationary distribution for both $\mathbf{v}^\mathsf{(HY)}$ and $\mathbf{v}^\mathsf{(EH)}$ is known, the performance of the proposed strategy with adaptive biases can be computed. First, the probability of power outage is a good measure of the battery level at EH-BSs and gives insight of the network general behavior, as we will show in Section \ref{sec:simulations}.
A power outage at the typical user U$_0$ occurs if it cannot find any base station to associate with, due to weak channel or power shortage at surrounding EH-BSs. Therefore, from the void probability of PPPs, the probability of power outage is expressed by:
\begin{align}
\mathbb{P}_{\text{out}} = \exp \left(- 
\Lambda_{\mathsf{EH}}^{(A)} \left( p_{\mathsf{L}}^{(\text{cov,EH})} \right)
\right)
\exp \left( - \sum_{X \in \left\lbrace \mathsf{HY} , \mathsf{OG} \right\rbrace}
\Lambda_{\mathsf{X}} \left( \mathsf{P}_\mathsf{Tx}^{(\max)}\right)
\right) 
\end{align}
where $\Lambda_{\mathsf{EH}}^{(A)}$ is the density of available EH-BSs and $\Lambda_{\mathsf{HY}}$ / $\Lambda_{\mathsf{OG}}$ is the total density of HY-/OG-BSs.
The first term  is the probability that no EH-BS is can be found within the maximum feasible power coverage $p_{\mathsf{L}}^{(\text{cov,EH})}$. The second term refers to the probability that none of the HY-BSs and OG-BSs satisfies the maximum transmit power constraint $\mathsf{P}_\mathsf{Tx}^{(\max)}$.
Note that the effect of the biases are implicitly captured in the first term only and is particularly noticeable when the network is poorly provided with access to the power grid (low densities $\lambda_{HY}$ and $\lambda_{OG}$). 

Next, we move onto the average power per area unit consumed from non-renewable sources, at both OG-BSs and HY-BSs.
Denoting $\mathbb{P}_{\text{T}}^{(\mathsf{G-OG})}(m)$ the probability that a OG-BS consumes $m$ units (computed in step 4 of the algorithm), we have:
\begin{align}
\mathsf{P}_{\mathsf{OG}}^{(G)} & =  \lambda_{\mathsf{OG}} \sum_{m \geq 1} m \; \mathbb{P}_{\text{T}}^{(\mathsf{G-OG})}(m) . 
\end{align}
Regarding HY-BSs, the probability that a HY-BS consumes $m$ units from the power grid depends also on its current battery level. A fully-charged HY-BS has much less probability to resort to the grid supply than HY-BSs with low battery level.
We denote $\mathbb{P}_{\text{T}}^{(\mathsf{G-HY})}(m \; \vert \; l)$  the probability that a HY-BS consumes $m$ units from the power grid given that $l$ units are stored in the battery. Recalling that $v_l^\mathsf{(HY)}$ is the probability to have $l$ power units in the battery, the average power per unit area consumed from the carbon-based grid supply at HY-BSs is equal to
\begin{align}
\mathsf{P}_{\mathsf{HY}}^{(G)} & = \lambda_{\mathsf{HY}} \underset{l=0}{\overset{\mathsf{L}}{\sum}} \sum_{m \geq 1}  m v_l^\mathsf{(HY)} \mathbb{P}_{\text{T}}^{(\mathsf{G-HY})}(m \; \vert \; l) .
\end{align}
Once again, the impact of $(\beta_A, \beta_G)$ is implicit and both $\mathbb{P}_{\text{OG}}^{(\mathsf{G})}$ and $\mathbb{P}_{\text{HY}}^{(\mathsf{G})}$ jointly depend on the stationary distribution of the battery states at EH-BSs and HY-BSs.

%
%
%
\section{Simulations and Performance results}
\label{sec:simulations}

In this Section, we validate the analysis proposed earlier and investigate the performance achieved by the proposed biased cell association, by comparison with conventional policies.

\subsection{Simulation set up and reference schemes}

In the following, the proposed policy is referred as "CA-A$\beta$", for cell association with adaptive biases. Without loss of generality, we can set $\beta_G$ to 1 and let $\beta_A$ vary. If not specified, we consider the simulation parameters of Table \ref{table:simulation_parameters}.
For performance comparison, we consider the three following reference schemes:
\begin{itemize}

	\item \emph{Conventional biased cell association (CA-F$\beta$):} association biases are fixed and depends on the BS type, regardless of the user required power or the BS current battery level as in \cite{Sakr2014,Song2014, Yang2016}
	. Biases are denoted as  $\beta_{\text{EH}}$, $\beta_{\text{HY}}$ and $\beta_{\text{OG}}$.
	
		\item \emph{Conventional non-biased cell association (CA-no$\beta$):} each user simply associates with the available BS that consumes the less transmit power (i.e. $\beta_{\text{X}}=1, \forall X$).
	
	\item \emph{Ideal scheme (Best-CA):} batteries are assumed always full. This reference gives an upper-bound on the achievable performance and is used to  efficiently compare a wide range of simulation parameters.
\end{itemize}

The proposed power-aware cell association takes advantage of (i) the power-availability criteria, (ii) the ability for users to choose their power supply and (iii) the adaptive biases.
The performance enhancement brought by the availability criteria has already been investigated in \cite{EHBS_Journal1}. To isolate its impact and fairly analyzing the gain provided by two other design parameters, we assume, for CA-F$\beta$, CA-no$\beta$ and Best-CA, that EH-BSs periodically broadcast their battery level, so that users determinate their power availability. 

\emph{Remark:} The performance obtained by Monte-Carlo simulations and by calculation is plotted in the two following figures. As observed, the plots show a good agreement between the computational and simulated results, which validates the approximation used to compute the density of served users and the algorithm of Section \ref{sec:markov_algo}.

\begin{table}
	\renewcommand{\arraystretch}{1.3}
	\caption{Simulation parameters}
	\label{table:simulation_parameters}
	\centering
	\begin{tabular}{|c|c||c|c||c|c|}
		\hline 
		\multicolumn{2}{|c||}{Network} &
		\multicolumn{2}{|c||}{PPP} &
		\multicolumn{2}{|c|}{Power} \\[3pt]
		\hline
		$\kappa$ / $\alpha$ & 1 / 4 &
		$\lambda_e$ / $N_e$ & 10\%$\mathsf{L}$ / 30 &
		$\mathsf{P_{Rx}}$ & -65dBm  \\[3pt]
		
		$\sigma$ & 4dB &
		$\omega$ & $\frac{50}{\pi 100^2}$ &
		$\mathsf{P_{Tx}^{(max)}}$ & 500mW \\[3pt]
		
		$\mathtt{A_{sim}}$ & 1km$^2$ & 
		$\beta_G$ & 1 &
		$\mathsf{L}$ / $\varepsilon \mathsf{L}$ & 1000 / 750mW  \\[3pt]
		\hline
	\end{tabular}
	\vspace{-10pt}
\end{table}

\subsection{On the probability of power outage}

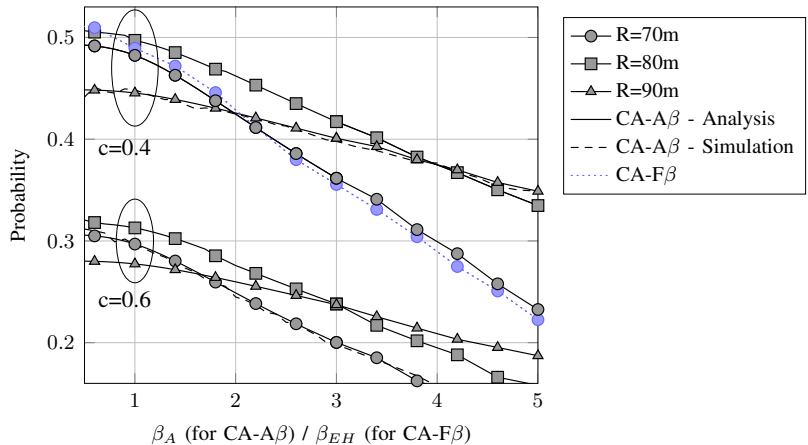
\begin{figure}
\centering
\centering \resizebox{0.65\columnwidth}{!}{%
\begin{tikzpicture}[scale=4,cap=round,>=latex]


  \begin{axis}[%
  ymin=0.16, ymax = 0.53,
  xmin=0.5, xmax = 5,
	xlabel= {$\beta_A$ (for CA-A$\beta$) / $\beta_{EH}$ (for CA-F$\beta$)},%
    ylabel= {Probability},%
    y label style={at={(axis description cs:-0.1,0.5)},rotate=0,anchor=south},
	every axis/.append style={font=\footnotesize},  
    grid=major,
   legend style={nodes=right,anchor=north west, at={(axis cs: 5.25,0.52)}},%
    mark size=2pt]

%
%

\addplot[solid,color=black,mark=*, mark options={solid,fill=black!40}, mark size=2.5pt, mark phase=2,  mark repeat=4, line width=0.5] table[x=betaA ,y=HY04 ,col sep=semicolon] {./Adapt_Outage_Ne_30_10L_P6550_OG_1000_HYEH_70.txt};
\addlegendentry{R=70m};

\addplot[solid,color=black,mark=square*, mark options={solid,fill=black!40}, mark size=2.5pt, mark phase=2,  mark repeat=4, line width=0.5] table[x=betaA ,y=HY04 ,col sep=semicolon] {./Adapt_Outage_Ne_30_10L_P6550_OG_1000_HYEH_80.txt};
\addlegendentry{R=80m};

\addplot[solid,color=black,mark=triangle*, mark options={solid,fill=black!40}, mark size=2.5pt, mark phase=2,  mark repeat=4, line width=0.5] table[x=betaA ,y=HY04 ,col sep=semicolon] {./Adapt_Outage_Ne_30_10L_P6550_OG_1000_HYEH_90.txt};
\addlegendentry{R=90m};

\addplot[solid,color=black, line width=0.5] table[x=betaA ,y=HY04,col sep=semicolon] {./Adapt_OutageBIS_Ne_30_10L_P6550_OG_1000_HYEH_80.txt};
\addlegendentry{CA-A$\beta$ - Analysis};

\addplot[dashed,color=black, line width=0.5] table[x=betaA ,y=HY04,col sep=semicolon] {./Adapt_Outage_Ne_30_10L_P6550_OG_1000_HYEH_90_REF.txt};
\addlegendentry{CA-A$\beta$ - Simulation};

\addplot[dotted,color=blue!60, line width=0.5] table[x=betaA ,y=HY04 ,col sep=semicolon] {./Fixed_Outage_Ne_30_10L_P6550_OG_1000_HYEH_70.txt};
\addlegendentry{CA-F$\beta$};

\addplot[dotted,color=blue!60,mark=*, mark options={solid,fill=blue!40}, mark size=2.5pt, line width=0.5] table[x=betaA ,y=HY04 ,col sep=semicolon] {./Fixed_Outage_Ne_30_10L_P6550_OG_1000_HYEH_70.txt};

\addplot[solid,color=black,mark=*, mark options={solid,fill=black!40}, mark size=2.5pt, mark phase=2,  mark repeat=4, line width=0.5] table[x=betaA ,y=HY04 ,col sep=semicolon] {./Adapt_Outage_Ne_30_10L_P6550_OG_1000_HYEH_70.txt};

\addplot[dashed,color=black, line width=0.5] table[x=betaA ,y=HY06,col sep=semicolon] {./Adapt_Outage_Ne_30_10L_P6550_OG_1000_HYEH_70_REF.txt};

\draw (axis cs: 1,0.47) ellipse (10pt and 25pt);
\draw [anchor=north] (axis cs: 0.9,0.41) node {\small c=0.4};

\addplot[solid,color=black,mark=*, mark options={solid,fill=black!40}, mark size=2.5pt, mark phase=2,  mark repeat=4, line width=0.5] table[x=betaA ,y=HY06 ,col sep=semicolon] {./Adapt_Outage_Ne_30_10L_P6550_OG_1000_HYEH_70.txt};

\addplot[solid,color=black,mark=square*, mark options={solid,fill=black!40}, mark size=2.5pt, mark phase=2,  mark repeat=4, line width=0.5] table[x=betaA ,y=HY06 ,col sep=semicolon] {./Adapt_Outage_Ne_30_10L_P6550_OG_1000_HYEH_80.txt};

\addplot[solid,color=black,mark=triangle*, mark options={solid,fill=black!40}, mark size=2.5pt, mark phase=2,  mark repeat=4, line width=0.5] table[x=betaA ,y=HY06 ,col sep=semicolon] {./Adapt_Outage_Ne_30_10L_P6550_OG_1000_HYEH_90.txt};

\draw (axis cs: 1,0.30) ellipse (8pt and 18pt);
\draw [anchor=north] (axis cs: 0.9,0.26) node {\small c=0.6};

\addplot[solid,color=black,mark=*, mark options={solid,fill=black!40}, mark size=2.5pt, line width=0.5] table[x=betaA ,y=HY04 ,col sep=semicolon] {./Adapt_OutageBIS_Ne_30_10L_P6550_OG_1000_HYEH_70.txt};
\addplot[solid,color=black,mark=*, mark options={solid,fill=black!40}, mark size=2.5pt, line width=0.5] table[x=betaA ,y=HY06 ,col sep=semicolon] {./Adapt_OutageBIS_Ne_30_10L_P6550_OG_1000_HYEH_70.txt};

\addplot[solid,color=black,mark=square*, mark options={solid,fill=black!40}, mark size=2.5pt, line width=0.5] table[x=betaA ,y=HY04 ,col sep=semicolon] {./Adapt_OutageBIS_Ne_30_10L_P6550_OG_1000_HYEH_80.txt};
\addplot[solid,color=black,mark=square*, mark options={solid,fill=black!40}, mark size=2.5pt, line width=0.5] table[x=betaA ,y=HY06 ,col sep=semicolon] {./Adapt_OutageBIS_Ne_30_10L_P6550_OG_1000_HYEH_80.txt};

\addplot[solid,color=black,mark=triangle*, mark options={solid,fill=black!40}, mark size=2.5pt, line width=0.5] table[x=betaA ,y=HY04 ,col sep=semicolon] {./Adapt_OutageBIS_Ne_30_10L_P6550_OG_1000_HYEH_90.txt};
\addplot[solid,color=black,mark=triangle*, mark options={solid,fill=black!40}, mark size=2.5pt, line width=0.5] table[x=betaA ,y=HY06 ,col sep=semicolon] {./Adapt_OutageBIS_Ne_30_10L_P6550_OG_1000_HYEH_90.txt};


    \end{axis}
\end{tikzpicture}
}
 \vspace{-10pt}
\caption{\footnotesize Loss in the overall power outage, with $\beta_G=1$, for a network consisting of c\% of HY-BS and (1-c)\% of EH-BS\vspace{-5pt}}
\label{fig:general_outage}
\end{figure}

We first consider a network where all small-cell BSs are provided with energy harvesting facilities and distributed according a PPP of density $\lambda = \frac{1}{\pi R^2}$, i.e. when BSs are R meters away one from each other in average. We assume that c\% of them are also connected to the power grid, i.e. $\lambda_{HY} = \frac{c}{\pi R^2}$, $\lambda_{EH} = \frac{1-c}{\pi R^2}$ and  $\lambda_{OG} = 0$.
Figure \ref{fig:general_outage} investigates, for different values of $R$, $c$ and $\beta_A$ (with $\beta_G=1$), the power outage probability achieved using biased association. 
It depicts the ratio of the power outage obtained with CA-A$\beta$ over the outage obtained in the ideal case Best-CA (with full batteries), i.e. $(\mathbb{P}_{\text{out}}^{(A\beta)} - \mathbb{P}_{\text{out}}^{(\textit{Best})} ) / \mathbb{P}_{\text{out}}^{(\textit{Best})} $. 
Given that the probability of power outage is dominated by the term $\exp \left( - \sum_{X \in \left\lbrace \mathsf{HY} , \mathsf{OG} \right\rbrace}
\Lambda_{\mathsf{X}} \left( \mathsf{P}_\mathsf{Tx}^{(\max)}\right)
\right) $, computing such ratio allows to isolate the gain issued from the different schemes and battery management.
As observed, the probability of power outage monotonically decreases with $\beta_A$, illustrating the fact that users are prompted to attach to HY-BSs with grid supply and that more energy is stored in the batteries without being consumed. The loss decreases till 0, i.e. batteries are always full and the outage probability cannot be further reduced.

As illustrated in Figure \ref{fig:general_outage}, both CA-A$\beta$ and CA-F$\beta$ provide fairly similar performance. Indeed, only users sufficiently far from HY-BSs may be in power outage and have thus to be served by the closest available EH-BS.  This suggests that the power outage probability is mostly impacted by the power-availability checking, performed prior to cell association, rather than by the adaptive biases. Thus, for the rest of this paper, we move onto investigating the on-grid power consumption.

\subsection{On the on-grid power consumption}

\subsubsection{First insights}

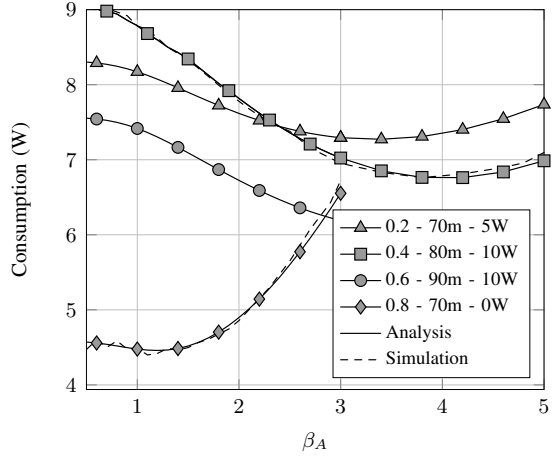
\begin{figure} 
	\centering
	\centering \resizebox{0.45\columnwidth}{!}{%
\begin{tikzpicture}[scale=4,cap=round,>=latex]


  \begin{axis}[%
   ymax = 9,
  xmin=0.5, xmax = 5,
	xlabel= {$\beta_A$},%
    ylabel= {Consumption (W)},%
    y label style={at={(axis description cs:-0.1,0.5)},rotate=0,anchor=south},
	every axis/.append style={font=\footnotesize},  
    grid=major,
   	legend style={nodes=right,font=\scriptsize},
    legend pos={south east},%
    mark size=2pt]

%
%

\addplot[solid,color=black,mark=triangle*, mark options={solid,fill=black!40}, mark size=3pt, mark phase=2, mark repeat=4, line width=0.5] table[x=betaA ,y=HY02_5,col sep=semicolon] {./Adapt_Ongrid_Ne_30_10L_P6550_OG_1000_HYEH_70.txt};
\addlegendentry{0.2 - 70m - 5W};
\addplot[solid,color=black,mark=square*, mark options={solid,fill=black!40}, mark size=2.5pt, mark phase=0, mark repeat=4, line width=0.5] table[x=betaA ,y=HY04_10,col sep=semicolon] {./Adapt_Ongrid_Ne_30_10L_P6550_OG_1000_HYEH_80.txt};
\addlegendentry{0.4 - 80m - 10W};
\addplot[solid,color=black, mark=*,mark options={solid,fill=black!40}, mark size=2.5pt, mark phase=2, mark repeat=4,line width=0.5] table[x=betaA ,y=HY06_10,col sep=semicolon] {./Adapt_Ongrid_Ne_30_10L_P6550_OG_1000_HYEH_90.txt};
\addlegendentry{0.6 - 90m - 10W};

\addplot[solid,color=black, mark=diamond*,mark options={solid,fill=black!40}, mark size=3pt, mark phase=2, mark repeat=4,line width=0.5] table[x=betaA ,y=HY08 ,col sep=semicolon] {./Adapt_Ongrid_Ne_30_10L_P6550_OG_1000_HYEH_70.txt};
\addlegendentry{0.8 - 70m - 0W};


\addplot[solid,color=black, line width=0.5] table[x=betaA ,y=HY04_10,col sep=semicolon] {./Adapt_Ongrid_Ne_30_10L_P6550_OG_1000_HYEH_80.txt};
\addlegendentry{Analysis};

\addplot[dashed,color=black, line width=0.5] table[x=betaA ,y=HY04_10,col sep=semicolon]
{./Adapt_Ongrid_Ne_30_10L_P6550_OG_1000_HYEH_REF.txt};
\addlegendentry{Simulation};

\addplot[dashed,color=black, line width=0.5] table[x=betaA ,y=HY08,col sep=semicolon]
{./Adapt_Ongrid_Ne_30_10L_P6550_OG_1000_HYEH_70_REF.txt};

\addplot[solid,color=black,mark=triangle*, mark options={solid,fill=black!40}, mark size=3pt, mark phase=1,  mark repeat=1, line width=0.5] table[x=betaA ,y=HY02_5,col sep=semicolon] {./Adapt_Ongrid_BISNe_30_10L_P6550_OG_1000_HYEH_70.txt};
\addplot[solid,color=black,mark=square*, mark options={solid,fill=black!40}, mark size=2.5pt, mark phase=1,  mark repeat=1, line width=0.5] table[x=betaA ,y=HY04_10,col sep=semicolon] {./Adapt_Ongrid_BISNe_30_10L_P6550_OG_1000_HYEH_80.txt};
%
%

%
%


    \end{axis}
\end{tikzpicture}
}
  \vspace{-10pt}
	\caption{\footnotesize Total on-grid power consumption over $\mathtt{A_{sim}}$, with $\lambda_{HY}=c \frac{1}{\pi R^2}$, $\beta_G=1$ and an offset  of n Watts - Legend: c, R, n. As the power consumption noticeably depends on the BS-user distance, we have subtracted a fixed offset to the obtained on-grid power consumption in Figure \ref{fig:general_ongrid}, for the sole purpose of clarity. This offset allows to plot in the same figure the performance achieved for a large range of BS densities.\vspace{-5pt}}
	\label{fig:general_ongrid}
\end{figure}

Figure \ref{fig:general_ongrid} depicts the total on-grid power consumed over the simulation area $\mathtt{A_{sim}}$ using CA-A$\beta$ and illustrates the balance between the stationary distribution of the battery level and the overall on-grid consumption. 
Increasing $\beta_A$ first allows to consume less power from the grid. For example, with $\lambda_{HY} = 40\% \frac{1}{\pi 80^2}$, more than 2W are saved by taking $\beta_A =4$ rather than $\beta_A \leq \beta_G = 1$. With a higher $\beta_A$, users located far from both HY-BSs and EH-BSs are prompted to select non-renewable energy for service even if, at a given time slot, sufficient renewable energy is stored in the battery of an EH-BS or a HY-BS. As a consequence, the battery level has higher probability to store more power units at next time slot, and thus, more users can be served from renewable energy.
Yet, passed a certain threshold, increasing  $\beta_A$ has a detrimental effect on the on-grid power consumption. Indeed, the battery level is kept uselessly high and users are let requesting too often grid power, given the channel conditions, power demand and BS density.

\subsubsection{Advantages of letting users decide on the power supply}

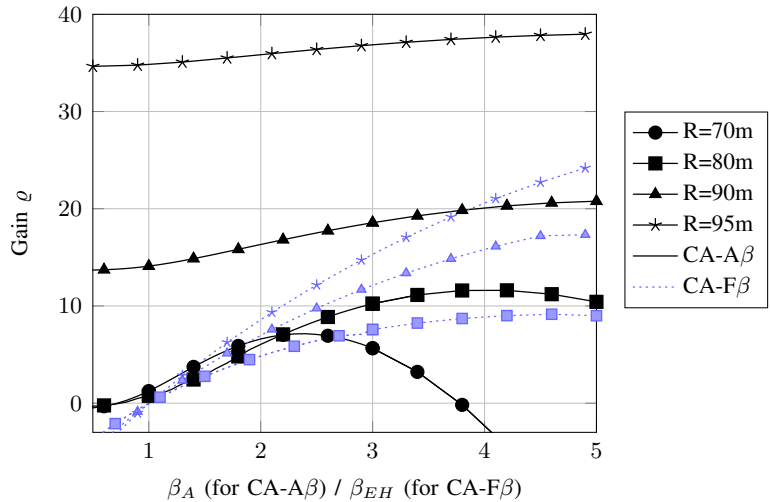
\begin{figure}
	\centering
	\centering \resizebox{0.62\columnwidth}{!}{%
\begin{tikzpicture}[scale=4,cap=round,>=latex]


  \begin{axis}[%
  ymin=-3, ymax = 40,
  xmin=0.5, xmax = 5,
	xlabel= {$\beta_A$ (for CA-A$\beta$) / $\beta_{EH}$ (for CA-F$\beta$)},%
    ylabel= {Gain $\varrho$},%
    y label style={at={(axis description cs:-0.1,0.5)},rotate=0,anchor=south},
	every axis/.append style={font=\footnotesize},  
    grid=major,
   legend style={nodes=right, at={(axis cs: 5.25, 30)},anchor=north west },
    mark size=2pt]

%
%
%

%
%
%
%


\addplot[solid,color=black, mark=*,mark size=2.5pt, mark options={solid}, mark phase=2, mark repeat=4,line width=0.5] table[x=betaA ,y=HY04 ,col sep=semicolon] {./Adapt_VarRho_court_Ne_30_10L_P6550_OG_1000_HYEH_70.txt};
\addlegendentry{R=70m};

\addplot[solid,color=black, mark=square*,mark size=2.5pt, mark options={solid}, mark phase=2, mark repeat=4,line width=0.5] table[x=betaA ,y=HY04 ,col sep=semicolon] {./Adapt_VarRho_court_Ne_30_10L_P6550_OG_1000_HYEH_80.txt};
\addlegendentry{R=80m};

\addplot[solid,color=black, mark=triangle*,mark size=2.5pt, mark options={solid}, mark phase=2, mark repeat=4,line width=0.5] table[x=betaA ,y=HY04 ,col sep=semicolon] {./Adapt_VarRho_Ne_30_10L_P6550_OG_1000_HYEH_90.txt};
\addlegendentry{R=90m};

\addplot[solid,color=black, mark=star,mark size=2.5pt, mark options={solid}, line width=0.5] table[x=betaA ,y=HY04 ,col sep=semicolon] {./Adapt_VarRho_Ne_30_10L_P6550_OG_1000_HYEH_95.txt};
\addlegendentry{R=95m};

\addplot[solid,color=black, line width=0.5] table[x=betaA ,y=HY04 ,col sep=semicolon] {./Adapt_VarRho_BISNe_30_10L_P6550_OG_1000_HYEH_70.txt};
\addlegendentry{CA-A$\beta$};
\addplot[dotted,color=blue!60, line width=0.5] table[x=betaA ,y=HY04 ,col sep=semicolon] {./Fixed_VarRho_Ne_30_10L_P6550_OG_1000_HYEH_80.txt};
\addlegendentry{CA-F$\beta$};

\addplot[solid,color=black, mark=*,mark size=2.5pt, mark options={solid}, line width=0.5] table[x=betaA ,y=HY04 ,col sep=semicolon] {./Adapt_VarRho_BISNe_30_10L_P6550_OG_1000_HYEH_70.txt};
\addplot[solid,color=black, mark=square*,mark size=2.5pt, mark options={solid}, line width=0.5] table[x=betaA ,y=HY04 ,col sep=semicolon] {./Adapt_VarRho_BISNe_30_10L_P6550_OG_1000_HYEH_80.txt};
\addplot[solid,color=black, mark=triangle*,mark size=2.5pt, mark options={solid},line width=0.5] table[x=betaA ,y=HY04 ,col sep=semicolon] {./Adapt_VarRho_BISNe_30_10L_P6550_OG_1000_HYEH_90.txt};



\addplot[dotted,color=blue!60, mark size=2pt, mark=square*,mark options={solid,fill=blue!40}, mark phase=1, mark repeat=2,line width=0.5] table[x=betaA ,y=HY04 ,col sep=semicolon] {./Fixed_VarRho_Ne_30_10L_P6550_OG_1000_HYEH_80.txt};

\addplot[dotted,color=blue!60, mark=triangle*,mark size=2pt, mark options={solid,fill=blue!40},line width=0.5] table[x=betaA ,y=HY04 ,col sep=semicolon] {./Fixed_VarRho_Ne_30_10L_P6550_OG_1000_HYEH_90.txt};

\addplot[dotted,color=blue!60,mark=square*,mark size=2pt, mark options={solid,fill=blue!40}, line width=0.5] table[x=betaA ,y=HY04 ,col sep=semicolon] {./Fixed_VarRho_BISNe_30_10L_P6550_OG_1000_HYEH_80.txt};

\addplot[dotted,color=blue!60, mark=star,mark size=2pt, mark options={solid,fill=blue!40},line width=0.5] table[x=betaA ,y=HY04 ,col sep=semicolon] {./Fixed_VarRho_Ne_30_10L_P6550_OG_1000_HYEH_95.txt};

    \end{axis}
\end{tikzpicture}
}
 \vspace{-10pt}
	\caption{\footnotesize  Gain $\varrho$ (\%) in the on-grid power consumption for a fixed proportion of HY-BSs, with $\lambda_{HY}= 0.4 \frac{1}{\pi R^2}$ and $\beta_G=1$ ($\beta_{HY}=1$).\vspace{-5pt}}
	\label{fig:varrho_HYEH_distance}
\end{figure}

Figure \ref{fig:varrho_HYEH_distance} further investigate such trade-off in a network deprived of OG-BSs, with varying overall density $\lambda= \frac{1}{\pi R^2}$ but always constituted of 40\% of HY-BSs and 60\% of EH-BSs.
We focus on the gain in the total on-grid power consumption $\mathsf{P}_{\mathsf{HY}}^{(\textit{A}\beta,G)}$ (resp. $\mathsf{P}_{\mathsf{HY}}^{(\textit{F}\beta,G)}$) provided by considering the biased strategy CA-A$\beta$ (resp. CA-F$\beta$), over the non-biased strategy CA-no$\beta$. It is computed as
\begin{align}
\varrho = \frac{\mathsf{P}_{\mathsf{HY}}^{(\textit{x}\beta,G)} - \mathsf{P}_{\mathsf{HY}}^{(\textit{No},G)} }{ \mathsf{P}_{\mathsf{HY}}^{(\textit{No},G)} - \mathsf{P}_{\mathsf{HY}}^{(\textit{Best},G)}}, \quad x = A \ (\text{resp. } F)
\end{align}
Note that the denominator includes $\mathsf{P}_{\mathsf{HY}}^{(\textit{Best},G)}$ as an offset to isolate the gain provided by biasing. The scaled consumption $\mathsf{P}_{\mathsf{HY}}^{(\textit{No},G)} - \mathsf{P}_{\mathsf{HY}}^{(\textit{Best},G)}$ can be understood as the maximum feasible range for performance improvement. 

We observe that, in denser networks (R$\leq$80m), most users are served from renewable energy sources whatever it be CA-A$\beta$, CA-F$\beta$ or CA-No$\beta$. The gain $\varrho$ achieved by biasing is thus limited. However, when the network density decreases, the available renewable power is not sufficient to manage all users in the network. Significant gain is obtained by considering biased cell associations as they prevent far users (requesting large amount of power) to completely deplete batteries and thereby, to deprive nearby users to access renewable supply in future time slots. For example with R=90m, 21\%-gain  is achieved by CA-A$\beta$ over CA-No$\beta$, which corresponds to around 9W over $\mathtt{A_{sim}}$.

In addition, the proposed CA-A$\beta$ outperforms the conventional CA-F$\beta$ by letting users decide on the power source at HY-BSs, implying that on-grid power may be consumed even if the battery is not empty. The resulted gain is particularly visible when $\beta_{EH} = \beta_{HY}=1$ and $\beta_A = \beta_G =1$, in which case $\varrho=0$ for CA-F$\beta$ and $\varrho>0$ for CA-A$\beta$.
Indeed, with the conventional CA-F$\beta$, batteries at HY-BSs are empty most of the time and strictly follow the fluctuation of the power unit arrivals. Each power unit is immediately consumed. This has three main drawbacks: (i) in case of power cut, HY-BSs become unavailable till sufficient energy is harvested, and the network must be handled by EH-BSs solely, which can cause severe data loss and long system recovery, (ii) little response is offered by HY-BSs to user traffic variations or increase in their power requirement, (iii) as all available power units are immediately consumed, the batteries at HY-BSs continuously suffer high level variations, which can drastically reduce their overall lifetime.
On the contrary, the proposed cell association with adaptive biases provides a similar battery management at both HY-BSs and EH-BSs, leads to smoother fluctuations of the battery level and allows a better power management in the time, from one slot to another.

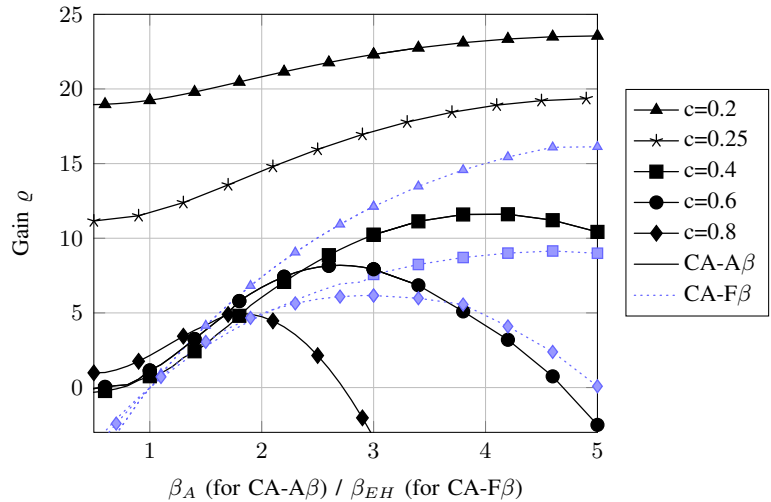
\begin{figure}
\centering
\centering \resizebox{0.62\columnwidth}{!}{%
\begin{tikzpicture}[scale=4,cap=round,>=latex]


  \begin{axis}[%
  ymin=-3, ymax = 25,
  xmin=0.5, xmax = 5,
	xlabel= {$\beta_A$ (for CA-A$\beta$) / $\beta_{EH}$ (for CA-F$\beta$)},%
    ylabel= {Gain $\varrho$},%
    y label style={at={(axis description cs:-0.1,0.5)},rotate=0,anchor=south},
	every axis/.append style={font=\footnotesize},  
    grid=major,
    legend style={nodes=right, at={(axis cs: 5.25, 20)},anchor=north west },
    mark size=2pt]

%
%

\addplot[solid,color=black, mark=triangle*,mark options={solid,fill=black}, mark size=2.5pt, mark phase=2,  mark repeat=4,line width=0.5] table[x=betaA ,y=HY02 ,col sep=semicolon] {./Adapt_VarRho_court_Ne_30_10L_P6550_OG_1000_HYEH_80.txt};
\addlegendentry{c=0.2};

\addplot[solid,color=black, mark=star,mark size=2.5pt, mark options={solid,fill=black}, line width=0.5] table[x=betaA ,y=HY025 ,col sep=semicolon] {./Adapt_VarRho_Ne_30_10L_P6550_OG_1000_HYEH_80_autres.txt};
\addlegendentry{c=0.25};
\addplot[solid,color=black,mark=square*, mark options={solid,fill=black}, mark size=2.5pt, mark phase=2,  mark repeat=4, line width=0.5] table[x=betaA ,y=HY04 ,col sep=semicolon] {./Adapt_VarRho_court_Ne_30_10L_P6550_OG_1000_HYEH_80.txt};
\addlegendentry{c=0.4};
\addplot[solid,color=black,mark=*, mark options={solid,fill=black}, mark size=2.5pt, mark phase=2,  mark repeat=4, line width=0.5] table[x=betaA ,y=HY06 ,col sep=semicolon] {./Adapt_VarRho_court_Ne_30_10L_P6550_OG_1000_HYEH_80.txt};
\addlegendentry{c=0.6};
\addplot[solid,color=black,mark=diamond*, mark options={solid,fill=black}, mark size=3pt, mark phase=1,  mark repeat=4, line width=0.5] table[x=betaA ,y=HY08 ,col sep=semicolon] {./Adapt_VarRho_court_Ne_30_10L_P6550_OG_1000_HYEH_80.txt};
\addlegendentry{c=0.8};

%
\addplot[solid,color=black, line width=0.5] table[x=betaA ,y=HY04 ,col sep=semicolon] {./Adapt_VarRho_BISNe_30_10L_P6550_OG_1000_HYEH_80.txt};
\addlegendentry{CA-A$\beta$};

\addplot[dotted,color=blue!60, line width=0.5] table[x=betaA ,y=HY02 ,col sep=semicolon] {./Fixed_VarRho_Ne_30_10L_P6550_OG_1000_HYEH_80.txt};
\addlegendentry{CA-F$\beta$};


\addplot[solid,color=black, mark=triangle*,mark size=2.5pt, mark options={solid,fill=black}, line width=0.5] table[x=betaA ,y=HY02 ,col sep=semicolon] {./Adapt_VarRho_BISNe_30_10L_P6550_OG_1000_HYEH_80.txt};

\addplot[solid,color=black, mark=square*,mark size=2.5pt, mark options={solid,fill=black}, line width=0.5] table[x=betaA ,y=HY04 ,col sep=semicolon] {./Adapt_VarRho_BISNe_30_10L_P6550_OG_1000_HYEH_80.txt};

\addplot[solid,color=black, mark=*,mark size=2.5pt, mark options={solid,fill=black}, line width=0.5] table[x=betaA ,y=HY06 ,col sep=semicolon] {./Adapt_VarRho_BISNe_30_10L_P6550_OG_1000_HYEH_80.txt};


\addplot[dotted,color=blue!60,mark=triangle*,mark size=2pt, mark options={solid,fill=blue!40}, mark phase=2,  mark repeat=2, line width=0.5] table[x=betaA ,y=HY02 ,col sep=semicolon] {./Fixed_VarRho_Ne_30_10L_P6550_OG_1000_HYEH_80.txt};

\addplot[dotted,color=blue!60,mark=diamond*,mark size=2.5pt, mark options={solid,fill=blue!40}, mark phase=1,  mark repeat=2, line width=0.5] table[x=betaA ,y=HY08 ,col sep=semicolon] {./Fixed_VarRho_Ne_30_10L_P6550_OG_1000_HYEH_80.txt};

\addplot[dotted,color=blue!60,mark=triangle*,mark size=2pt, mark options={solid,fill=blue!40}, line width=0.5] table[x=betaA ,y=HY02 ,col sep=semicolon] {./Fixed_VarRho_BISNe_30_10L_P6550_OG_1000_HYEH_80.txt};

\addplot[dotted,color=blue!60,mark=diamond*,mark size=2.5pt, mark options={solid,fill=blue!40}, line width=0.5] table[x=betaA ,y=HY08 ,col sep=semicolon] {./Fixed_VarRho_BISNe_30_10L_P6550_OG_1000_HYEH_80.txt};

\addplot[dotted,color=blue!60,mark size=2pt, mark options={solid,fill=blue!40}, mark phase=2,  mark repeat=2, line width=0.5] table[x=betaA ,y=HY04 ,col sep=semicolon] {./Fixed_VarRho_Ne_30_10L_P6550_OG_1000_HYEH_80.txt};
\addplot[dotted,color=blue!60,mark=square*,mark size=2pt, mark options={solid,fill=blue!40}, line width=0.5] table[x=betaA ,y=HY04 ,col sep=semicolon] {./Fixed_VarRho_BISNe_30_10L_P6550_OG_1000_HYEH_80.txt};


\addplot[solid,color=black,mark=*, mark options={solid,fill=black}, mark size=2.5pt, mark phase=2,  mark repeat=4, line width=0.5] table[x=betaA ,y=HY06 ,col sep=semicolon] {./Adapt_VarRho_court_Ne_30_10L_P6550_OG_1000_HYEH_80.txt};

    \end{axis}
\end{tikzpicture}
}
 \vspace{-10pt}
\caption{\footnotesize Gain $\varrho$ (\%) in the total on-grid power consumption over \textit{No}-$\beta$, with $\lambda_{HY}= \frac{c}{\pi 80^2}$ and $\beta_G=1$ ($\beta_{HY}=1$). \vspace{-5pt}}
\label{fig:varrho_HYEH_pourcentage}
\end{figure}

\subsubsection{Advantages of the proposed adaptive gains}

In Figure \ref{fig:varrho_HYEH_pourcentage}, we consider a network with fixed density $\lambda= \frac{1}{\pi 80^2}$ but with varying proportion of HY-BSs. We note that the case R=80m has been chosen on purpose since CA-A$\beta$ and CA-F$\beta$ achieve similar performance gain for c=0.4, as illustrated in Figure \ref{fig:varrho_HYEH_distance}.
First, when most of base stations are hybrid (e.g. c=80\%), the gain $\varrho$ remains rather limited for CA-A$\beta$ (resp. CA-F$\beta$) and is achieved for $\beta_A \rightarrow \beta_G = 1$ (resp. $\beta_{EH} \rightarrow \beta_{HY} = 1$), implying that distinguishing the type of BS powering for cell association does not provide much power gain. The conventional CA-F$\beta$ slightly outperforms the proposed CA-A$\beta$ in this case. Given the high homogeneity of such network, preventing batteries to be empty as does the proposed strategy, tends to increase the on-grid power consumption. Yet, this result does not account for the three main drawbacks detailed earlier.

On the contrary, when the proportion of HY-BSs is decreasing, the proposed cell association significantly outperforms the non-biased strategy CA-No$\beta$ and more than 23\% of gain is obtained when only 20\% of base stations are HY-BSs (c=0.2). In this case, the optimal $\beta_A$ is such that $\beta_A \gg \beta_G$ and increases when the proportion of HY-BSs diminishes, thus preventing far users to request renewable energy, even if feasible at a given instant. Moreover, distinguishing cell association based on the type of powering used for data transmission (as in CA-A$\beta$) rather than on the type of BS (as in CA-F$\beta$) significantly improves the network performance. While 23\% of gain is obtained with the proposed strategy, only a 16\%-gain is reached with conventional biased cell association, which represents an additional gain of 3W (Figure \ref{fig:general_ongrid}).
Indeed, the proposed CA-A$\beta$ provides a better distribution of the power available throughout the network, not only among EH-BSs but also among HY-BSs.
In addition, the range of $\beta_A$ for which high power gain is achieved is much larger than for CA-F$\beta$. Thus, the proposed cell association still provides satisfying performance gain even if the set $(\beta_A, \beta_G)$ is away from its optimum. This trend is also observed in Figure \ref{fig:varrho_HYEH_distance}, for lower network densities.

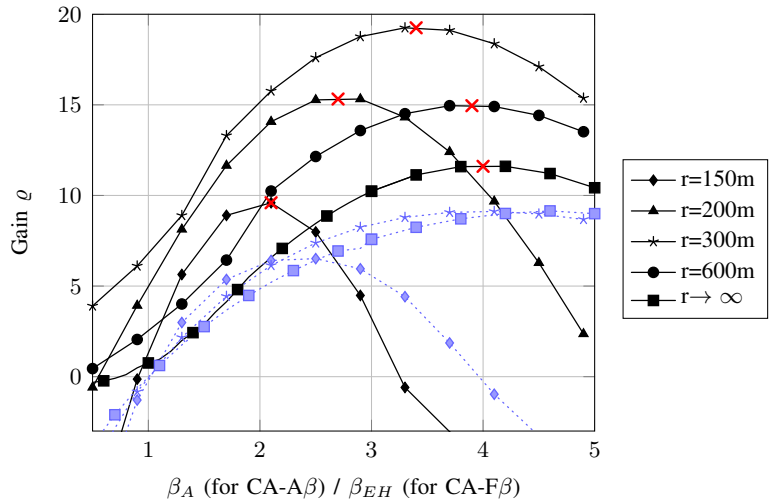
\begin{figure}
\centering
\centering \resizebox{0.62\columnwidth}{!}{%
\begin{tikzpicture}[scale=4,cap=round,>=latex]


  \begin{axis}[%
  ymin=-3, ymax = 20,
  xmin=0.5, xmax = 5,
	xlabel= {$\beta_A$ (for CA-A$\beta$) / $\beta_{EH}$ (for CA-F$\beta$)},%
    ylabel= {Gain $\varrho$},%
    y label style={at={(axis description cs:-0.1,0.5)},rotate=0,anchor=south},
	every axis/.append style={font=\footnotesize},  
    grid=major,
    legend style={nodes=right, at={(axis cs: 5.25,12)}, anchor= north west},%
    mark size=2pt]

%
%

\addplot[solid,color=black, mark=diamond*,mark size=2pt, mark options={solid,fill=black}, line width=0.5] table[x=betaA ,y=HY0150 ,col sep=semicolon] {./Adapt_VarRho_Ne_30_10L_P6550_OG_vary_HYEH_80.txt};
\addlegendentry{r=150m};

\addplot[solid,color=black, mark=triangle*,mark size=2pt, mark options={solid,fill=black}, line width=0.5] table[x=betaA ,y=HY0200 ,col sep=semicolon] {./Adapt_VarRho_Ne_30_10L_P6550_OG_vary_HYEH_80.txt};
\addlegendentry{r=200m};
\addplot[solid,color=black,mark=star, mark options={solid,fill=black}, mark size=2pt, line width=0.5] table[x=betaA ,y=HY0300 ,col sep=semicolon] {./Adapt_VarRho_Ne_30_10L_P6550_OG_vary_HYEH_80.txt};
\addlegendentry{r=300m};

\addplot[solid,color=black,mark=*, mark options={solid,fill=black}, mark size=2pt,  line width=0.5] table[x=betaA ,y=HY0600 ,col sep=semicolon] {./Adapt_VarRho_Ne_30_10L_P6550_OG_vary_HYEH_80.txt};
\addlegendentry{r=600m};
%
%
\addplot[solid,color=black,mark=square*, mark options={solid,fill=black}, mark size=2pt, mark phase=2,  mark repeat=4, line width=0.5] table[x=betaA ,y=HY04 ,col sep=semicolon] {./Adapt_VarRho_court_Ne_30_10L_P6550_OG_1000_HYEH_80.txt};
\addlegendentry{r$\rightarrow \infty$};

%

%
\addplot[solid,color=black, mark=square*,mark size=2pt, mark options={solid,fill=black}, line width=0.5] table[x=betaA ,y=HY04 ,col sep=semicolon] {./Adapt_VarRho_BISNe_30_10L_P6550_OG_1000_HYEH_80.txt};
%
%
%


\addplot[dotted,color=blue!60, mark size=2pt, mark=square*,mark options={solid,fill=blue!40}, mark phase=1, mark repeat=2,line width=0.5] table[x=betaA ,y=HY04 ,col sep=semicolon] {./Fixed_VarRho_Ne_30_10L_P6550_OG_1000_HYEH_80.txt};
\addplot[dotted,color=blue!60,mark=square*,mark size=2pt, mark options={solid,fill=blue!40}, line width=0.5] table[x=betaA ,y=HY04 ,col sep=semicolon] {./Fixed_VarRho_BISNe_30_10L_P6550_OG_1000_HYEH_80.txt};

\addplot[dotted,color=blue!60, mark size=2pt, mark=diamond*,mark options={solid,fill=blue!40}, line width=0.5] table[x=betaA ,y=HY0150 ,col sep=semicolon] {./Fixed_VarRho_Ne_30_10L_P6550_OG_vary_HYEH_80.txt};

\addplot[dotted,color=blue!60, mark size=2pt, mark=star,mark options={solid,fill=blue!40}, line width=0.5] table[x=betaA ,y=HY0300 ,col sep=semicolon] {./Fixed_VarRho_Ne_30_10L_P6550_OG_vary_HYEH_80.txt};

\addplot[only marks,color=red, mark size=3pt, mark=x,mark options={solid,fill=red}, line width=1] coordinates
{(2.100000, 9.597391)};

\addplot[only marks,color=red, mark size=3pt, mark=x,mark options={solid,fill=red}, line width=1] coordinates
{(2.7,15.320542)};

\addplot[only marks,color=red, mark size=3pt, mark=x,mark options={solid,fill=red}, line width=1] coordinates
{(3.4, 19.250972)};

\addplot[only marks,color=red, mark size=3pt, mark=x,mark options={solid,fill=red}, line width=1] coordinates
{(3.9,14.953528)};

\addplot[only marks,color=red, mark size=3pt, mark=x,mark options={solid,fill=red}, line width=1] coordinates
{(4,11.607296)};


    \end{axis}
\end{tikzpicture}
}
 \vspace{-10pt}
\caption{\footnotesize Gain (\%) in the total on-grid power consumption over \textit{No}-$\beta$, with $\lambda_{HY}= \frac{0.4}{\pi 80^2}$, $\lambda_{OG}= \frac{1}{\pi r^2}$ and $\beta_G=1$ ($\beta_{HY}=1$ / optimal $\beta_{OG}$). \vspace{-5pt}}
\label{fig:varrho_HYEH_OGvary}
\end{figure}

\subsubsection{Performance in highly heterogeneous networks}

As last result, we analyze the impact of OG-BSs and assume as basis a network of density $\lambda= \frac{1}{\pi 80^2}$ and constituted of 40\% of HY-BSs and 60\% of EH-BSs. To this network, we progressively add OG-BSs, by varying their density $\lambda_{OG}= \frac{1}{\pi r^2}$. The case $r \rightarrow \infty$ refers to the case analyzed earlier, i.e. without OG-BSs.
As illustrated in Figure \ref{fig:varrho_HYEH_OGvary}, the conventional biased strategy CA-F$\beta$ does not benefit from additional OG-BSs. The gain $\varrho$ remains quite constant for any $r$. As their batteries are most of the time empty, HY-BSs tend to behave like OG-BSs and the network is equivalent, in terms of user association, to having only EH-BSs and OG-BSs/HY-BSs, with respective densities $\frac{0.6}{\pi 80^2}$ and $\lambda_{OG} + \frac{0.4}{\pi 80^2}$. This boils down to the case analyzed in Figure \ref{fig:varrho_HYEH_distance}.

On the contrary, the proposed cell association is significantly impacted by additional OG-BSs, as explained in the following. The lower the density of OG-BSs is, the higher should be the bias $\beta_A$ to maximize $\varrho$. Yet, the variation of the maximal gain $\varrho$  (illustrated by red crosses) as a function of $r$ is concave and upper-bounded by the case r=300m. 
The increasing part ($r\leq 300$m) is in line with what has been observed in Figures \ref{fig:varrho_HYEH_distance} and \ref{fig:varrho_HYEH_pourcentage}. Little gain is obtained in networks consisting in majority of BSs connected to the grid supply. 
When the overall network density is decreasing, the distance between a user and its serving BS is longer, more transmit power is required and increasing $\beta_A$ to prompt users to request grid supply maintains higher battery level and thus, allows a better energy distribution among BSs. Yet, when $r\geq 300$m, the density of OG-BSs is too low to affect performance results. The power gain $\varrho$ starts to decrease and tends to the case $r \rightarrow \infty$.
Given such observation, we can conclude that the proposed cell association largely takes advantage of the high heterogeneity of the networks.

\section{Conclusion}
\label{sec:conclusion}

We have addressed the issue of the heterogeneity in the base station powering and have proposed a novel user association with adaptive biases. The  provided  gain is three-fold. First, users do not partition BSs depending on their effective type, but depending on the power supply, renewable or not, that is used for data transmission. Second, users associated with a HY-BS are let free to decide on the power supply to be used, and may request power from the grid even if the battery is not empty.
Third, the power coverage of BSs as perceived by users is controlled by adaptive biases, which are set at each user and at each time slot, depending on the current BS battery level, the power required to satisfy a received power constraint and the estimated power consumed to serve other users potentially associated with the same BS. 
We have shown that this novel strategy significantly outperforms conventional policies, particularly in power-constrained networks, with low density or with limited access to the grid power supply. Moreover, it allows a better distribution of the available renewable energy among base stations and thereby, takes advantage of higher heterogeneity in the BS powering.

\appendix 

We briefly re-state the closed-form expressions obtained in \cite{EHBS_Journal1} and used in step 4 of the algorithm.
Given $\Omega_{\text{X}}^{(\mathcal{A})} (p \; \vert \; l) $ in Eq. \eqref{eq:Omega_A},
the probability $ \mathbb{P}_{\text{T}}^{(\mathsf{A-X})}(m \; \vert \; l)$ that a X-BS consumes exactly $m$ power units from the battery is computed recursively as:

\begin{align}
&\mathbb{P}_{\text{T}}^{(\mathsf{A-X})} (m \; \vert \; l) = \left \lbrace
\begin{array}{ll}
\frac{\mathbb{P}_{\Sigma} \left(m,\Omega_{\text{X}}^{(\mathcal{A})}(p \; \vert \; l),\lfloor p_{l}^{(\text{cov})} \rfloor\right) }{ \mathbb{P}_{\text{sum}} (l) } & \; \text{if} \; m \leq l \\
0 & \; \text{otherwise}.
\end{array}\right.
\label{eq:prob_T_l}
\\
&\text{where } \; \mathbb{P}_{\text{sum}} (l) = \sum_{m=1}^{l} \mathbb{P}_{\Sigma} \left(m,\Omega_{\text{X}}^{(\mathcal{A})}(p \; \vert \; l),\lfloor p_{l}^{(\text{cov})} \rfloor\right)
\nonumber 
\\
&\text{and } \;
 \left \lbrace \begin{array}{rl}
\mathbb{P}_{\Sigma} \left(0,\Omega,P\right) 
&=\exp \left(-\Omega (P) \right) 
\\
\mathbb{P}_{\Sigma} \left(m,\Omega_,P\right)&= \underset{q=1}{\overset{m^{\star}}{\sum}}  \frac{q}{m} \; C_q \;  \mathbb{P}_{\Sigma}(m-q,\Omega,P)
\end{array} \right.
\nonumber 
\end{align}

The probability $\mathbb{P}_{\text{T}}^{(\mathsf{G-X})}(m \; \vert \; l)$ to consume $m$ units from the power grid is computed in a similar manner, using  $\Omega_{\text{X}}^{(\mathcal{A})} (p \; \vert \; l) $ of Eq. \eqref{eq:Omega_G}. 

Finally, the transition matrix $\mathbf{P}^\mathsf{(X)} = \left[ \mathbb{P}_{\overrightarrow{lq}}^\mathsf{(X)}\right]$ is given by
\begin{align}
\forall l 
\left \lbrace
\begin{array}{ll}
\mathbb{P}_{\overrightarrow{lq}} =
\sum_m \mathbb{P}_{\text{T}}^{(\mathsf{A-X})} \left( m \; \vert \; l \right) \mathbb{P}_{\text{H}}^{(X)}\left( q-l+m \right)
& \quad \forall q \neq \mathsf{L} 
\\
\mathbb{P}_{\overrightarrow{l \mathsf{L} }} =
\sum_m \mathbb{P}_{\text{T}}^{(\mathsf{A-X})} \left( m \; \vert \; l \right) \sum_{q \geq \mathsf{L}-l+m} \mathbb{P}_{\text{H}}^{(X)}\left(q \right)
&
\end{array}
\right.
\label{eq:P_ij}
\end{align}
with $\mathbb{P}_{\text{H}}^{(X)}\left(m \right)$ the probability that a X-BS harvests $m$ power units during a time slot.

\fontsize{10}{10}
\selectfont

\bibliographystyle{IEEEtranN}
{\footnotesize 
\bibliography{biblio_EHBS_journal2}
}

\end{document}